\documentclass[aip,jcp,reprint]{revtex4-2}

\usepackage{amsmath}
\usepackage{amssymb}
\usepackage{bm}
\usepackage{graphicx}
\usepackage{hyperref}
\usepackage{mathptmx}
\usepackage{xcolor}

\usepackage[utf8]{inputenc}
\usepackage[T1]{fontenc}

\makeatletter
\def\@email#1#2{
\endgroup
\patchcmd{\titleblock@produce}
{\frontmatter@RRAPformat}
{\frontmatter@RRAPformat{\produce@RRAP{*#1\href{mailto:#2}{#2}}}\frontmatter@RRAPformat}
{}{}
}
\makeatother

\begin{document}

\title{The WEST code for large-scale excited-state materials simulations}

\author{Victor Wen-zhe Yu}
\affiliation{Materials Science Division, Argonne National Laboratory, Lemont, Illinois 60439, United States}

\author{Siyuan Chen}
\affiliation{Pritzker School of Molecular Engineering, The University of Chicago, Chicago, Illinois 60637, United States}

\author{Yu Jin}
\affiliation{Pritzker School of Molecular Engineering, The University of Chicago, Chicago, Illinois 60637, United States}
\affiliation{Initiative for Computational Catalysis, Flatiron Institute, New York, New York 10010, United States}

\author{Vrindaa Somjit}
\affiliation{Materials Science Division, Argonne National Laboratory, Lemont, Illinois 60439, United States}

\author{Stefano Paolo Villani}
\affiliation{Pritzker School of Molecular Engineering, The University of Chicago, Chicago, Illinois 60637, United States}

\author{Jiawei Zhan}
\affiliation{Pritzker School of Molecular Engineering, The University of Chicago, Chicago, Illinois 60637, United States}

\author{Marco Govoni}
\affiliation{Department of Physics, Computer Science, and Mathematics, University of Modena and Reggio Emilia, Modena, 41125, Italy}

\author{Giulia Galli}
\thanks{Corresponding author, e-mail:~gagalli@uchicago.edu}
\affiliation{Pritzker School of Molecular Engineering, The University of Chicago, Chicago, Illinois 60637, United States}
\affiliation{Department of Chemistry, The University of Chicago, Chicago, Illinois 60637, United States}
\affiliation{Materials Science Division, Argonne National Laboratory, Lemont, Illinois 60439, United States}

\date{\today}

\begin{abstract}
We present WEST, an open-source plane-wave pseudopotential code for large-scale excited-state materials simulations, and describe its theoretical foundations, software architecture, and capabilities. WEST implements full-frequency GW, quantum defect embedding theory, the Bethe-Salpeter equation, and time-dependent density functional theory within a common algorithmic framework that avoids the explicit computation of virtual electronic states. By combining density functional and density matrix perturbation theory, low-rank representations of the dielectric screening and exact exchange, and localization techniques, WEST achieves favorable computational scaling with system size. The code supports the calculation of quasi-particle and neutral excitation energies, optical and photoluminescence spectra, excited-state forces, and non-adiabatic couplings, with interoperable workflows connecting to quantum chemistry, vibronic coupling, and quantum computing packages. A hierarchical parallelization strategy and graphics processing unit (GPU) acceleration deliver near-ideal strong scaling to thousands of GPUs, enabling accurate excited-state simulations of systems with more than a thousand atoms. Representative applications, spanning the full optical cycle of solid-state spin defects, self-trapped excitons in metal-halide perovskites, and the optical response of liquid water and ice, demonstrate the accuracy and versatility of the code across diverse material classes. The capabilities implemented in WEST establish the code as a scalable platform for predictive excited-state simulations, high-throughput materials discovery, and the generation of high-fidelity datasets for machine learning in computational materials science.
\end{abstract}

\maketitle

\section{Introduction}

First-principles simulations based on Kohn-Sham density functional theory (KS-DFT)~\cite{dft_hohenberg_1964,dft_kohn_1965} have become indispensable tools for understanding and predicting the ground-state properties of systems ranging from molecules to materials. However, many phenomena of central importance in areas such as microelectronics, optics, catalysis, and quantum technologies are governed by excited-state processes that are not accurately described within DFT, thus requiring more advanced theoretical approaches~\cite{perspective_galli_2026}. Time-dependent DFT (TDDFT)~\cite{tddft_marques_2004,tddft_casida_2012} has emerged as a widely used approach for neutral excitations because of its favorable balance between accuracy and computational cost. In addition, methods based on many-body perturbation theory (MBPT), including the GW approximation (G and W stand for the single-particle Green's function and the screened Coulomb interaction, respectively)~\cite{gw_hybertsen_1985,gw_hybertsen_1986,gw_strinati_1988,gw_aryasetiawan_1998} for charged excitations and the Bethe-Salpeter equation (BSE)~\cite{bse_salpeter_1951,bse_albrecht_1998,bse_onida_2002} for neutral excitations, provide rigorous frameworks for the computation of excited-state properties. Recently, several quantum embedding approaches~\cite{embedding_sun_2016,embedding_jones_2020,quantum_vorwerk_2022} have been developed to describe the electronic structure of different regions within a complex system at distinct levels of theory, enabling the treatment of localized, multi-configurational states using correlated wave function methods that would be prohibitively expensive if applied to the entire system.

Despite the successes of these methods, their application to materials in realistic environments remains computationally challenging. Conventional implementations of TDDFT and MBPT rely on explicit calculations of a large number of empty electronic states, and the associated computational cost scales unfavorably with system size. Moreover, most implementations lack an efficient computation of excited-state forces, necessary to investigate excited-state structural relaxations and emission processes. Overcoming these challenges is essential for the simulation of complex materials in experimentally relevant environments, as material properties are often dictated by structural inhomogeneities such as dislocations, surfaces, interfaces, and nanostructures, whose description requires supercells containing hundreds to thousands of atoms.

High-performance computing (HPC) is now in the exascale era, dominated by massive graphics processing unit (GPU) acceleration of codes, calling for electronic structure software to be specifically designed to efficiently exploit such platforms. Furthermore, the rapid advancement of artificial intelligence and machine learning (AI/ML) in materials science has created an urgent demand for large, high-fidelity datasets of excited-state properties that conventional implementations cannot affordably generate at scale.

The open-source WEST code~\cite{west_govoni_2015,west_website} has been developed to address these challenges through scalable formulations of excited-state electronic-structure methods that avoid explicit computations of empty electronic states and large dielectric matrices. WEST provides a unified software framework for large-scale excited-state simulations based on MBPT, TDDFT, and quantum embedding. Core methodologies implemented in WEST include full-frequency G$_0$W$_0$ calculations~\cite{west_govoni_2015,soc_scherpelz_2016,gw100_govoni_2018,west_yu_2022} using the projective dielectric eigenpotential (PDEP) technique~\cite{pdep_nguyen_2012,pdep_pham_2013}, BSE and TDDFT formulations based on density matrix perturbation theory (DMPT)~\cite{bse_rocca_2010,bse_rocca_2012,bse_ping_2012,tddft_jin_2022,tddft_jin_2023,bse_yu_2024}, and quantum defect embedding theory (QDET) for systems with multi-configurational states~\cite{qdet_sheng_2022,qdet_chen_2025}. Beyond quasi-particle (QP) energies, vertical excitation energies, and optical absorption spectra, WEST enables the computation of analytical excited-state forces~\cite{tddft_jin_2023}, non-adiabatic coupling (NAC)~\cite{nac_villani_2026}, and a broad range of quantities through interoperable external software packages and automated workflows.

Defining features of WEST are its scalability and applicability to large systems. Using algorithms that eliminate explicit summations over empty states makes the convergence of calculations with respect to computational parameters straightforward and systematic. Computational cost and memory usage are further reduced through low-rank approximations and localization techniques, complemented by a hierarchical parallelization strategy that achieves excellent scalability on HPC architectures based on central processing units (CPUs) and GPUs, thereby making it feasible to perform accurate excited-state simulations of systems containing more than one thousand atoms~\cite{west_yu_2022,bse_yu_2024,dislocation_zhang_2026}.

In this paper, we present a comprehensive overview of the scientific scope of the WEST code and its capabilities for large-scale excited-state materials simulations. In section~\ref{sec:features}, we review the theoretical methods and algorithms implemented in the code, including G$_0$W$_0$, QDET, BSE, and TDDFT. In section~\ref{sec:technical}, we discuss the software design, parallelization strategy, GPU acceleration, and interoperability that underpin the performance and flexibility of WEST. Section~\ref{sec:applications} highlights the capabilities of the code through representative applications involving complex excited-state phenomena across a broad range of materials. Finally, our concluding remarks and perspective for future developments are given in section~\ref{sec:summary}.

\section{Functionalities of the WEST code}
\label{sec:features}

As mentioned in the Introduction, starting from ground-state DFT (equation~\ref{eq:ks}) calculations performed within the plane-wave pseudopotential framework, WEST enables excited-state calculations using a broad range of methods, including full-frequency G$_0$W$_0$ (equations~\ref{eq:qp}--\ref{eq:pdep}), QDET (equation~\ref{eq:qdet}), BSE, and TDDFT (equations~\ref{eq:bse_tddft}--\ref{eq:nac2}). These capabilities are complemented by an extensive set of postprocessing tools for the analysis and interpretation of both ground- and excited-state properties (equations~\ref{eq:rho}--\ref{eq:tdm}).

The WEST code is designed to be applicable to both finite and periodic systems, with support for spin-polarized calculations. All code functionalities support both semilocal and hybrid functionals, including PBE0 and HSE06, as well as the dielectric-dependent hybrid (DDH)~\cite{ddh_skone_2014,ddh_skone_2016} and screened-exchange range-separated hybrid (SE-RSH)~\cite{sersh_zhan_2023,sersh_zhan_2025} functionals. In DDH and SE-RSH, the fraction of exact exchange is determined non-empirically from the dielectric screening of the material. In SE-RSH, both the fraction of the exact exchange and the range separation parameter are spatially dependent rather than global constants, and are derived from first principles, thus leading to an accurate description of heterogeneous materials with varying local dielectric environments. Through the use of the adaptively compressed exchange (ACE) method~\cite{ace_lin_2016}, hybrid functional calculations can be performed at a computational cost comparable to semilocal functionals~\cite{tddft_jin_2023,hybrid_yu_2025}. All WEST functionalities support scalar-relativistic norm-conserving pseudopotentials, while G$_0$W$_0$ and TDDFT additionally support fully relativistic norm-conserving pseudopotentials including spin-orbit coupling (SOC) effects for a rigorous treatment of systems containing heavy elements~\cite{soc_scherpelz_2016}.

\begin{figure}[ht!]
\includegraphics[width=0.48\textwidth]{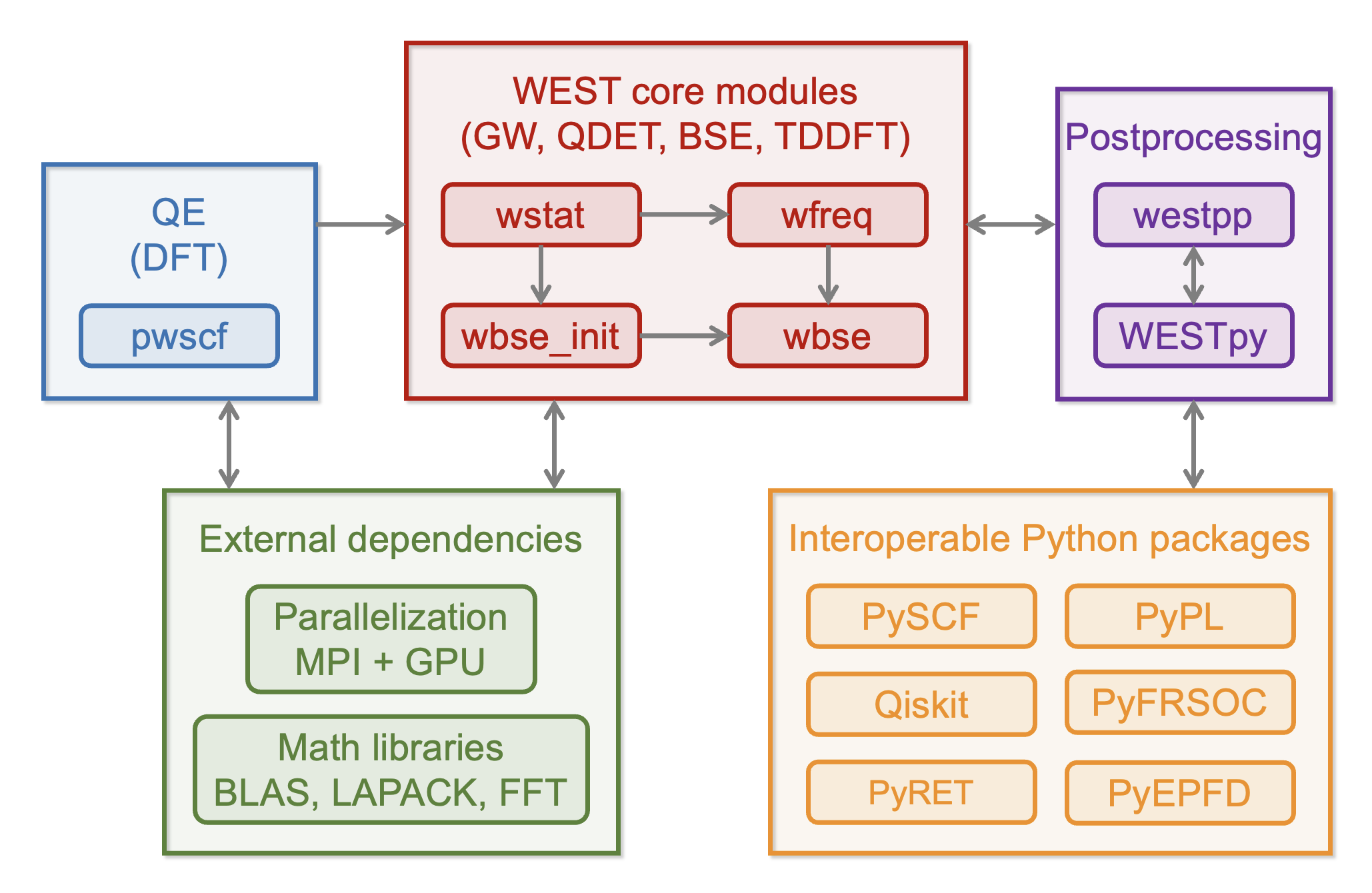}
\caption{Schematic overview of the general workflow of the WEST code. The \texttt{pwscf} code of Quantum ESPRESSO (QE) is used for ground-state DFT calculations. The core modules of WEST include \texttt{wstat} and \texttt{wfreq} for G$_0$W$_0$ and QDET calculations, and \texttt{wbse\_init} and \texttt{wbse} for BSE and TDDFT calculations. Postprocessing of WEST calculations is carried out with \texttt{westpp} and \texttt{WESTpy}. WEST depends on standard scientific computing libraries and is interfaced with multiple Python packages for the computation of a broad range of quantities.}
\label{fig:workflow}
\end{figure}

A schematic overview of the general workflow of WEST is shown in figure~\ref{fig:workflow}. Ground-state DFT calculations (equation~\ref{eq:ks}) are first carried out using the \texttt{pwscf} code of the Quantum ESPRESSO software package~\cite{qe_giannozzi_2020}. The KS wave functions and energies are then used as input for the core modules of WEST, including \texttt{wstat} for the diagonalization of the static dielectric screening (equation~\ref{eq:pdep}), \texttt{wfreq} for the computation of full-frequency G$_0$W$_0$ QP energies (equation~\ref{eq:eqp}) or parameters of the QDET effective Hamiltonian (equation~\ref{eq:qdet}), \texttt{wbse\_init} for the evaluation of bare or screened Coulomb integrals (equations~\ref{eq:tau_u} and \ref{eq:tau}), and \texttt{wbse} for the diagonalization of the BSE or TDDFT Liouville superoperator (equation~\ref{eq:bse_tddft}), the computation of excited-state forces (equation~\ref{eq:forces}), and the evaluation of NACs (equations~\ref{eq:nac1} and \ref{eq:nac2}). Postprocessing tools \texttt{westpp} and the \texttt{WESTpy} Python package enable further data analysis, visualization, and automated workflows (equations~\ref{eq:rho}--\ref{eq:tdm}). Finally, WEST interfaces with multiple Python packages for the computation of additional quantities and physical observables (section~\ref{sec:interop}).

In the following, we summarize the main functionalities of WEST, focusing on electronic and optical properties of materials. Detailed methodological and algorithmic descriptions are provided in the referenced literature.

\subsection{Full-frequency G$_0$W$_0$ calculations}
\label{sec:gw}

In KS-DFT~\cite{dft_hohenberg_1964,dft_kohn_1965}, the ground state of a system of interacting electrons in the external field of the ions is obtained by solving a set of single-particle equations,
\begin{equation}
\label{eq:ks}
H_{\mathrm{KS}} \psi_i = \varepsilon_i \psi_i \,,
\end{equation}
where $\psi_i$ and $\varepsilon_i$ are the wave function and energy of the $i$-th KS state, respectively. The KS Hamiltonian, $H_{\mathrm{KS}}$, includes the single-particle kinetic energy operator and the Hartree, external (ionic), and XC potential operators. QP states are obtained by solving a similar set of equations,
\begin{equation}
\label{eq:qp}
H_{\mathrm{QP}} \psi_i^{\mathrm{QP}} = \varepsilon_i^{\mathrm{QP}} \psi_i^{\mathrm{QP}} \,,
\end{equation}
where the QP Hamiltonian, $H_{\mathrm{QP}}$, is defined by replacing, in $H_{\mathrm{KS}}$, the XC potential $v_{\mathrm{xc}}$ with the nonlocal, frequency-dependent self-energy operator $\Sigma$, which, in the G$_0$W$_0$ approximation~\cite{gw_hybertsen_1985,gw_hybertsen_1986,gw_aryasetiawan_1998}, is expressed as
\begin{equation}
\label{eq:sigma}
\Sigma (\mathbf{r}, \mathbf{r}'; \omega) = i \int_{- \infty}^{+ \infty} \frac{\mathrm{d} \omega'}{2 \pi} G_0 (\mathbf{r}, \mathbf{r}'; \omega + \omega') W_0 (\mathbf{r}, \mathbf{r}'; \omega') \,.
\end{equation}
Within the full-frequency G$_0$W$_0$ method, the frequency integration is carried out numerically with the contour deformation technique~\cite{contour_godby_1988,contour_lebegue_2003} without relying on simplified plasmon-pole models.

The terms $G_0$ and $W_0$ are computed non-self-consistently from the KS wave functions and energies obtained by solving equation~\ref{eq:ks} with a chosen XC functional. QP energies are computed using perturbation theory starting from the solution of equation~\ref{eq:ks},
\begin{align}
\varepsilon_i^{\mathrm{QP}} & = \varepsilon_i + \langle \psi_i | H_{\mathrm{QP}} - H_{\mathrm{KS}} | \psi_i \rangle \nonumber \\
& = \varepsilon_i + \langle \psi_i | \Sigma (\varepsilon_i^{\mathrm{QP}}) - v_{\mathrm{xc}} | \psi_i \rangle \,,
\label{eq:eqp}
\end{align}
enabling predictions of fundamental band gaps, ionization potentials, and electron affinities.

We partition $W$ into two parts, $W = v + W_p$, where $v$ is the bare Coulomb potential, and the frequency-dependent part $W_p$ is related to the density-density response function $\chi$ as $W_p = v \chi v = v^{1/2} \bar{\chi} v^{1/2}$, with $\bar{\chi}$ being the symmetrized counterpart of $\chi$. To compute $\bar{\chi}$ at zero frequency without explicitly computing empty states or inverting large dielectric matrices, we use the PDEP technique~\cite{dielectric_wilson_2008,dielectric_wilson_2009,pdep_nguyen_2012,pdep_pham_2013}, which iteratively diagonalizes the symmetrized irreducible density-density response function $\bar{\chi}^0$. Using the relation $\bar{\chi} = \bar{\chi}^0 + \bar{\chi}^0 \bar{\chi}$ in the random phase approximation (RPA), $\bar{\chi}$ can be expressed as
\begin{equation}
\label{eq:pdep}
\bar{\chi} = \Xi + \frac{1}{\Omega} \sum_a^{N_{\mathrm{PDEP}}} | \bar{\varphi}_a \rangle \frac{\lambda_a}{1 - \lambda_a} \langle \bar{\varphi}_a | \,,
\end{equation}
where $\lambda_a$ and $\bar{\varphi}_a$ denote the $a$-th eigenvalue and eigenfunction of $\bar{\chi}^0$, respectively; the $\Xi$ term takes into account the long-range dielectric response, $\Omega$ is the volume of the simulation cell, and $N_{\mathrm{PDEP}}$ is the number of eigen-potentials that determines the accuracy of the low-rank approximation in equation~\ref{eq:pdep}. The density response is evaluated within density functional perturbation theory (DFPT)~\cite{dfpt_baroni_1987,dfpt_baroni_2001} by solving the Sternheimer equation~\cite{sternheimer_sternheimer_1954}. The evaluation of the self-energy at multiple frequencies is then carried out with an iterative algorithm based on Lanczos chains~\cite{gw_umari_2010,pdep_nguyen_2012}. This formulation eliminates the need for separate convergence parameters associated with the number of empty states and any energy cutoff for response functions. The computational complexity of the algorithm scales as $\mathcal{O}(N^4)$, or more specifically, $N_{\mathrm{occ}}^2 \times N_{\mathrm{PDEP}} \times N_{\mathrm{pw}}$, while that of the conventional Adler-Wiser formulation~\cite{dielectric_adler_1962,dielectric_wiser_1963} scales as $N_{\mathrm{occ}} \times N_{\mathrm{empty}} \times N_{\mathrm{pw}}^2$, where $N$ is the number of electrons, $N_{\mathrm{occ}}$ ($N_{\mathrm{empty}}$) is the number of occupied (empty) states, and $N_{\mathrm{pw}}$ is the number of plane-waves used to represent the wave functions (in practice $N_{\mathrm{pw}} \gg N_{\mathrm{PDEP}}$). It is worth noting that the strategy of avoiding summations over empty states in the computation of dielectric screening is not restricted to plane-waves and is also applicable to localized, nonorthogonal basis sets, as demonstrated in reference~\onlinecite{localized_hubener_2012}. However, its computational advantage is generally more pronounced in plane-wave calculations, where the number of empty states can become very large and constitute a significant computational bottleneck.

The full-frequency G$_0$W$_0$ implementation of WEST has been extensively benchmarked and validated on molecular GW100~\cite{gw100_govoni_2018} and mixed molecular/material GW-SOC81~\cite{soc_scherpelz_2016} datasets. In addition, WEST and Qbox can be coupled to carry out G$_0$W$_0$ calculations beyond the RPA, using a finite-field approach that includes XC effects in the dielectric response~\cite{gw_ma_2019}.

\subsection{Quantum defect embedding theory (QDET)}
\label{sec:qdet}

QDET is a Green's function-based quantum embedding method to describe multi-configurational electronic states~\cite{qdet_sheng_2022,qdet_chen_2025}. The central idea of QDET is to partition the single-particle states of the entire system into an active space and an environment. The former consists of a small number of localized states, for example, those associated with a point defect in semiconductors or insulators, while the latter includes all remaining states of the host solid. The electronic structure of the active space is described using accurate many-body quantum chemistry methods, which are required to capture multi-configurational characters. The environment is treated at the G$_0$W$_0$ level of theory. This partitioning allows one to achieve high accuracy in the description of localized and correlated states at a computational cost significantly lower than that of a full many-body treatment of the entire system.

Within QDET, the electronic structure of the full system is mapped onto an effective Hamiltonian of the form
\begin{equation}
\label{eq:qdet}
H^{\mathrm{eff}} = \sum_{ij \in A} t^{\mathrm{eff}}_{ij} c^\dagger_i c_j + \frac{1}{2} \sum_{ijkl \in A} v^{\mathrm{eff}}_{ijkl} c^\dagger_i c^\dagger_j c_l c_k \,,
\end{equation}
where the one-body term $t^{\mathrm{eff}}$ incorporates the KS Hamiltonian in the active space while removing double-counted terms in the high- and low-level descriptions of the system. The two-body term $v^{\mathrm{eff}}$ captures the effective interaction between the active space and the environment through a static, partially screened Coulomb potential. For systems with strong coupling between the active space and the environment, a hybridization term is included in $H^{\mathrm{eff}}$, defined on the space spanned by the active space orbitals and a set of auxiliary bath orbitals~\cite{qdet_chen_2025}. The effective or auxiliary Hamiltonian is frequency-independent and can be diagonalized using an impurity solver such as full-configuration interaction (FCI), selected configuration interaction (SCI), multi-reference configuration interaction singles and doubles (MR-CISD), phaseless auxiliary field quantum Monte Carlo (ph-AFQMC), or specialized algorithms on quantum computers~\cite{quantum_huang_2022,quantum_vorwerk_2022,quantum_huang_2023}. The resulting many-body wave functions enable computations of Green's functions and spectral functions, inter-system crossing (ISC)~\cite{isc_jin_2025} matrix elements, and non-resonant energy transfer processes~\cite{ret_chattaraj_2024,ret_chattaraj_2025}.

The construction of the QDET Hamiltonian relies on the G$_0$W$_0$ implementation described in the previous section to evaluate off-diagonal matrix elements of the self-energy, with a scaling of $\mathcal{O}(N^3 M^2)$, where $M$ is the number of orbitals in the active space. The computational cost of diagonalizing the Hamiltonian depends on the size of the active space and the choice of the impurity solver.

\subsection{Bethe-Salpeter equation (BSE) and time-dependent density functional theory (TDDFT)}
\label{sec:bse_tddft}

\subsubsection{Vertical excitation energies and optical absorption spectra}

In WEST, the BSE and TDDFT implementations avoid explicit summations over empty states by reformulating the problem within DMPT~\cite{bse_rocca_2010,bse_rocca_2012,bse_ping_2012,tddft_jin_2023,bse_yu_2024}. This approach eliminates the need to construct large Hamiltonians in an electron-hole basis and enables efficient calculations for large systems. Within the Tamm-Dancoff approximation (TDA)~\cite{tda_hirata_1999}, the vertical excitation energy (VEE) $\omega_s$ from the ground to the $s$-th excited state is obtained by solving an eigenvalue problem,
\begin{equation}
\label{eq:bse_tddft}
(D + K^{1e} - K^{1d}) A_s = \omega_s A_s \,,
\end{equation}
where $(D + K^{1e} - K^{1d})$ is the Liouville superoperator, with the diagonal term $D$ accounting for the single-particle energy difference between electrons and holes, and the exchange term $K^{1e}$ and direct term $K^{1d}$ describing the electron-hole interactions. The TDDFT and BSE formulations differ only in the specific definitions of the $D$, $K^{1e}$, and $K^{1d}$ operators, and two complementary iterative algorithms are implemented in WEST for TDDFT and BSE. The Davidson algorithm is used to diagonalize the Liouville superoperator, obtaining, for a selected number of excited states, the VEEs $\omega_s$ and the auxiliary orbitals $A_s = \{ | a_{s,v} \rangle: v = 1, \dots, N_{\mathrm{occ}} \}$ that enter the definition of the linear change of the density matrix with respect to the ground state, due to the $s$-th neutral excitation:
\begin{equation}
\Delta \gamma_s = \sum_v^{N_{\mathrm{occ}}} | a_{s,v} \rangle \langle \psi_v | \,.
\end{equation}
Both spin-conserving and spin-flip excitations are supported~\cite{tddft_jin_2023,bse_yu_2024}. The Lanczos algorithm allows for the direct computation of optical absorption spectra over a broad energy range without explicitly computing $\omega_s$ and $A_s$~\cite{lanczos_walker_2006,lanczos_rocca_2008}.

In the hybrid TDDFT case, the calculation of the $K^{1e}$ term requires the evaluation of bare Coulomb integrals of the form
\begin{equation}
\label{eq:tau_u}
\tau_{vv'}^u (\mathbf{r}) = \int \mathrm{d} \mathbf{r}' v_c (\mathbf{r}, \mathbf{r}') \psi_v (\mathbf{r}') \psi_{v'}^* (\mathbf{r}') \,,
\end{equation}
where $v_c (\mathbf{r}, \mathbf{r'}) = (|\mathbf{r} - \mathbf{r'}|)^{-1}$ is the bare Coulomb potential. In the BSE case, one needs instead to compute screened Coulomb integrals
\begin{equation}
\label{eq:tau}
\tau_{vv'} (\mathbf{r}) = \int \mathrm{d} \mathbf{r}' W (\mathbf{r}, \mathbf{r}') \psi_v (\mathbf{r}') \psi_{v'}^* (\mathbf{r}')
\end{equation}
for all pairs of occupied KS wave functions. The computational cost of evaluating these integrals is significantly reduced by transforming the occupied wave functions into a set of localized Wannier functions,
\begin{equation}
\label{eq:wannier}
\tilde{\psi}_v = \sum_{v'}^{N_{\mathrm{occ}}} U_{vv'} \psi_{v'} \,,
\end{equation}
where the unitary transformation $U$ is determined via the joint approximate diagonalization of eigen-matrices (JADE) algorithm~\cite{wannier_gygi_2003,jade_holobar_2006}. In WEST, the GPU-accelerated implementation of JADE supports the efficient localization of electronic states for systems with thousands of occupied orbitals. Coulomb integrals are evaluated only for pairs of overlapping Wannier functions, reducing the number of integrals by more than an order of magnitude~\cite{bse_nguyen_2019,tddft_jin_2023,bse_yu_2024,bse_elliott_2019,bse_marsili_2017}, relative to those between KS orbitals.

The computational cost of the WEST TDDFT implementation scales as $\mathcal{O}(N^3)$. Compared to the conventional BSE formulation with a formal scaling of $\mathcal{O}(N^6)$, the WEST BSE implementation has a significantly reduced computational complexity. The overall scaling is $\mathcal{O}(N^4)$, where the evaluation of the screened Coulomb interaction using the PDEP technique scales as $\mathcal{O}(N^4)$~\cite{west_yu_2022,west_govoni_2015,dielectric_wilson_2008,dielectric_wilson_2009,pdep_nguyen_2012,pdep_pham_2013}, and the application of the Liouville superoperator scales as $\mathcal{O}(N^3)$~\cite{bse_nguyen_2019,bse_marsili_2017}. This reduced scaling extends the applicability of BSE to large systems~\cite{bse_yu_2024}. In addition, WEST can be coupled with the Qbox DFT code to perform BSE calculations beyond the RPA using a finite-field approach~\cite{bse_nguyen_2019}.

\subsubsection{Analytical excited-state nuclear forces}

Analytical excited-state nuclear forces within the TDDFT~\cite{tddft_jin_2023} and BSE~\cite{bseforces_jin_2026} frameworks are implemented in WEST using the extended Lagrangian formalism~\cite{forces_hutter_2003}, supporting both spin-conserving and spin-flip excited states as well as the use of hybrid functionals. The forces of the $s$-th excited state are computed as
\begin{equation}
\label{eq:forces}
F_{s,I\alpha} = F_{\mathrm{GS},I\alpha} - \frac{\mathrm{d} \omega_s}{\mathrm{d} R_{I\alpha}},
\end{equation}
where $F_{\mathrm{GS},I\alpha}$ denotes the ground-state forces acting on the $I$-th atom along the $\alpha$-th Cartesian direction computed from DFT calculations. The derivative of the VEE $\omega_s$ is evaluated as
\begin{equation}
\label{eq:derivative}
\frac{\mathrm{d} \omega_s}{\mathrm{d} R_{I\alpha}} = \int \mathrm{d} \mathbf{r} \frac{\partial V_{\mathrm{ext} (\mathbf{r})}}{\partial R_{I\alpha}} \left[ \Delta \rho_{s}^{(x)} (\mathbf{r}) + \Delta \rho_{s}^{(z)} (\mathbf{r}) \right],
\end{equation}
where $\frac{\partial V_{\mathrm{ext} (\mathbf{r})}}{\partial R_{I\alpha}}$ is the partial derivative of the external potential described using pseudopotentials, $ \Delta \rho_{s}^{(x)} (\mathbf{r})$ is the unrelaxed differential density
\begin{equation}
\begin{aligned}
\label{eq:diff_rho}
\Delta \rho_s^{(x)} (\mathbf{r}) = & \sum_v^{N_{\mathrm{occ}}} |a_{s,v} (\mathbf{r})|^2 \\
& - \sum_v^{N_{\mathrm{occ}}} \sum_{v'}^{N_{\mathrm{occ}}} \psi_v (\mathbf{r}) \psi_{v'}^* (\mathbf{r}) \int \mathrm{d} \mathbf{r'} a_{s,v}^* (\mathbf{r'}) a_{s,v'} (\mathbf{r'}) \,,
\end{aligned}
\end{equation}
and
\begin{equation}
\Delta \rho_s^{(z)} (\mathbf{r}) = \sum_v^{N_{\mathrm{occ}}} \left[ Z_v^* (\mathbf{r}) \psi_v (\mathbf{r}) + \psi_v^* (\mathbf{r}) Z_v (\mathbf{r}) \right]
\end{equation}
accounts for the orbital relaxation effects in the excitation process. Here, the vectors $|Z_v\rangle$ are computed by solving the Handy-Schaefer $Z$-vector equation~\cite{forces_handy_1984}, whose solution is accelerated by combining the ACE method~\cite{ace_lin_2016}, Wannier localization of KS wave functions~\cite{wannier_gygi_2003}, and the inexact Krylov subspace technique~\cite{ikrylov_simoncini_2003,ikrylov_vandenesof_2004} in a scalable GPU implementation~\cite{tddft_jin_2023}. The computation of analytical forces exhibits the same scaling as VEE calculations, making it applicable to large systems with hundreds to thousands of atoms.

\subsubsection{Non-adiabatic coupling (NAC)}

Many excited-state phenomena, including ultrafast photochemical and photophysical processes, charge and energy transfer, and non-radiative decays, are governed by non-adiabatic processes, i.e., non-radiative spin-conserving transitions between electronic states induced by the nuclear motion. Therefore, the description of the coupled electron–nuclear dynamics beyond the Born–Oppenheimer approximation is essential~\cite{nonadiabatic_tully_1990,nonadiabatic_curchod_2013,nonadiabatic_curchod_2018}, especially near conical intersections where two adiabatic potential energy surfaces intersect~\cite{intersect_herzberg_1963,intersect_yarkony_2001,intersect_matsika_2011,intersect_matsika_2021}. Accurate treatment of such processes requires the calculation of derivative couplings induced by nuclear motion, motivating the implementation of analytical NAC vectors in the hybrid TDDFT framework of WEST~\cite{nac_villani_2026}. The NAC vectors are defined as the derivative coupling between two electronic many-body states with respect to the displacement $R_{I\alpha}$ of atom $I$ in direction $\alpha$, namely
\begin{align}
d_{0s,I\alpha} & = \big\langle 0 \big| \frac{\mathrm{d} s}{\mathrm{d} R_{I\alpha}} \big\rangle \,, \label{eq:nac1} \\
d_{s's,I\alpha} & = \big\langle s' \big| \frac{\mathrm{d} s}{\mathrm{d} R_{I\alpha}} \big\rangle \,, \label{eq:nac2}
\end{align}
where $|0\rangle$ is the wave function of the ground state, $|s\rangle$ $(|s'\rangle)$ represents the wave function of the $s$-th ($s'$-th) excited state, and the excitations can be spin-conserving or spin-flip. The implementation in WEST builds upon a generalization of the $Z$-vector method~\cite{nac_wang_2021,nac_liu_2023} employed for the forces to express the total derivatives in equations~\ref{eq:nac1} and \ref{eq:nac2} in terms of the auxiliary orbitals $\{a_{s,v}\}$ and the KS wave functions $\{\psi_v\}$ analogously to equation~\ref{eq:derivative}. The implementation leverages the same optimized, scalable GPU code of TDDFT forces, with acceleration strategies including the ACE method, Wannier localization of KS wave functions, and the inexact Krylov subspace technique for solving the $Z$-vector equations. It has been exploited to compute electron-phonon (e-ph) matrix elements between excited states at the hybrid TDDFT level to describe internal conversion (IC) processes in spin defects~\cite{ic_villani_2026}. The capability to compute excited-state energies, forces, and NACs efficiently and at the same level of theory constitutes a crucial step towards non-adiabatic molecular dynamics in extended and heterogeneous systems.

\subsection{Postprocessing functionalities}
\label{sec:pp}

WEST provides a comprehensive suite of postprocessing tools for analyzing ground- and excited-state properties, as summarized below.

\subsubsection{Ground-state electronic structure}

\begin{itemize}
\item Visualization of KS wave functions $\{\psi_i (\mathbf{r})\}$ and the electron density,
\begin{equation}
\label{eq:rho}
\rho (\mathbf{r}) = \sum_v^{N_{\mathrm{occ}}} |\psi_v (\mathbf{r})|^2 \,.
\end{equation}
Volumetric data can be exported in the Gaussian cube format. Planar averages along $xy$, $yz$, or $zx$ planes and spherical averages can be computed.

\item Computation and plotting of density of states (DOS),
\begin{equation}
\mathrm{DOS}(E) = \sum_i \delta(E - \varepsilon_i) \,,
\end{equation}
where $\delta$ is the Dirac delta function approximated by a Gaussian broadening, and local density of states (LDOS) along a specific direction (e.g., $z$) of the simulation cell,
\begin{equation}
\label{eq:ldos}
\mathrm{LDOS}(z,E) = \sum_i \int \frac{\mathrm{d}x}{L_x} \int \frac{\mathrm{d}y}{L_y} | \psi_i (x,y,z) | ^2 \delta(E - \varepsilon_i) \,,
\end{equation}
where $L_x$ and $L_y$ are the lengths of the $x$ and $y$ axes of the simulation cell, respectively.

\item Computation of transition dipole moments between KS wave functions,
\begin{equation}
\mathbf{\mu}_{ij} = \langle \psi_i | \mathbf{r} | \psi_j \rangle = \frac{\langle \psi_i | [H_{\mathrm{KS}}, \mathbf{r}] | \psi_j \rangle}{\varepsilon_i - \varepsilon_j} \,.
\end{equation}

\item Computation of localization factor (LF) and inverse participation ratio (IPR) of KS wave functions, for the identification of localized states originating from, e.g., defects in solids.
\begin{align}
\mathrm{LF}_i & = \int_{\Omega_s} \mathrm{d}^3 \mathbf{r} |\psi_i (\mathbf{r})|^2 \,, \\
\mathrm{IPR}_i & = \int \mathrm{d}^3 \mathbf{r} |\psi_i (\mathbf{r})|^4 \,,
\end{align}
where the integration of the LF is carried out within a sphere $\Omega_s$, and the integration of the IPR is carried out in the entire simulation cell.

\item Computation of the unitary transformation matrix $U$ that localizes KS wave functions (equation~\ref{eq:wannier}), and the associated Wannier centers.
\end{itemize}

\subsubsection{G$_0$W$_0$ and QDET results}

\begin{itemize}
\item Visualization of PDEP functions $\{\bar{\varphi}_a\}$ in the cube format.

\item Analysis and plotting of frequency dependence of the self-energy, which can be used to assess the validity of solving equation~\ref{eq:qp} by linearization, and diagnose the issue of multiple roots.

\item Diagonalization of QDET Hamiltonians using FCI or SCI as implemented in PySCF. By specifying the point group of the system, including its symmetry operations and character table, a symmetry analysis can be performed to determine the irreducible representations of the KS single-particle states and QDET many-body states.
\end{itemize}

\subsubsection{BSE/TDDFT excited states}

\begin{itemize}
\item Visualization of the transition density
\begin{equation}
\Delta \rho_s (\mathbf{r}) = \sum_v^{N_{\mathrm{occ}}} a_{s,v} (\mathbf{r}) \psi_v^* (\mathbf{r}) \,,
\end{equation}
and the unrelaxed differential density (equation~\ref{eq:diff_rho}).

\item Decomposition of excited states into electron-hole pairs,
\begin{equation}
\label{eq:decomp}
c_{s,vc} = \langle a_{s,v} | \psi_c \rangle \,,
\end{equation}
where $v = 1, \dots, N_{\mathrm{occ}}$ and $c = N_{\mathrm{occ}} + 1, \dots, N_{\mathrm{occ}} + N_{\mathrm{empty}}$.

\item Computation of the transition dipole moment
\begin{equation}
\label{eq:tdm}
\mu_s = \int \mathrm{d} \mathbf{r} \Delta \rho_{s} (\mathbf{r}) \mathbf{r} = \sum_v^{N_{\mathrm{occ}}} \langle \psi_v | \mathbf{r} | a_{s,v} \rangle \,,
\end{equation}
in either real space or reciprocal space.

\item Computation of spin multiplicity $\langle S^2 \rangle$.

\item Computation and plotting of optical absorption spectra.

\item Excited-state geometry relaxation using BSE/TDDFT forces (equation~\ref{eq:forces}) and the Broyden-Fletcher-Goldfarb-Shanno (BFGS) optimization algorithm, implemented in an automated workflow that, at each BFGS step, runs the ground-state DFT calculation and the excited-state BSE/TDDFT calculation, extracts the excited-state energies and forces, and advances the atomic positions until convergence is reached.
\end{itemize}

\section{Technical aspects of the WEST code}
\label{sec:technical}

The WEST code is designed to meet the computational demands of large-scale excited-state calculations with a strong emphasis on combining accuracy, efficiency, and scalability. In this section, we describe the key technical components that underpin the performance and versatility of the code. We begin with its multi-level parallelization strategy, which exposes the full extent of algorithmic parallelism and enables the efficient utilization of distributed-memory computers. We then discuss GPU acceleration, highlighting the design choices leading to high performance on GPUs while maintaining a unified code base. Finally, we outline the software engineering practices, interoperability features, and ecosystem integrations that support the adoption and extensibility of WEST within the broader computational materials science community.

\subsection{Parallelization}

In the WEST code, parallelism is exposed at multiple layers. At the highest level, eigenvectors of the operator to be diagonalized, e.g., $\bar{\varphi}_a$ in equation~\ref{eq:pdep} or $A_s$ in equation~\ref{eq:bse_tddft}, are distributed, enabling embarrassingly parallel execution of key steps in many-body perturbation theory and TDDFT calculations. Additional levels of parallelism distribute spin channels, bands, plane-wave coefficients, and fast Fourier transform (FFT) grid points using the message passing interface (MPI). Achieving high performance in large-scale simulations requires an appropriate combination of all parallelization levels, with the optimal configuration depending on both the simulation setup for any specific system and the hardware. In practice, the overall performance is often dominated by the efficiency of FFTs. On CPUs, FFTs are most efficient when carried out on a single socket or node to reduce costly all-to-all communication in parallel FFTs. For all parallelization levels, data communication is carefully organized to allow for overlap of communication and computation and minimize synchronization overhead. The WEST hierarchical parallelization enables fine-grained workload balancing and efficient utilization of distributed-memory HPC resources. The scalability of WEST has been demonstrated on leadership-class supercomputers utilizing up to 500,000 CPU cores for the simulation of a solid/liquid interface~\cite{west_govoni_2015}.

\subsection{GPU acceleration}

WEST supports GPU acceleration through a directive-based programming model, OpenACC, to offload compute-intensive kernels to GPU accelerators, with ongoing efforts towards supporting OpenMP target offloading for improved portability across varied hardware platforms. The use of OpenACC enables a single-source code design, where computational kernels are annotated with directives and offloaded with minimal modifications to the CPU code base, thereby preserving maintainability and portability across heterogeneous architectures.

High performance on GPUs is achieved through several complementary strategies~\cite{west_yu_2022,tddft_jin_2023,bse_yu_2024}. First, WEST leverages vendor-optimized libraries for key numerical operations, such as cuFFT and cuBLAS for FFT and linear algebra, respectively. Second, the GPU implementation retains the same hierarchical parallelization scheme as the CPU version, which helps minimize expensive data movement between GPUs by ensuring that FFTs are executed either on a single GPU or on the minimum number of GPUs with sufficient device memory to host the simulation data. Third, non-blocking MPI communications are used to overlap inter-GPU communications with computations, further improving parallel efficiency and achieving excellent scalability to 25,920 GPUs~\cite{west_yu_2022}. In addition, selected parts of the code can utilize mixed-precision arithmetic to further enhance performance without compromising numerical accuracy.

\begin{figure*}[ht!]
\includegraphics[width=0.72\textwidth]{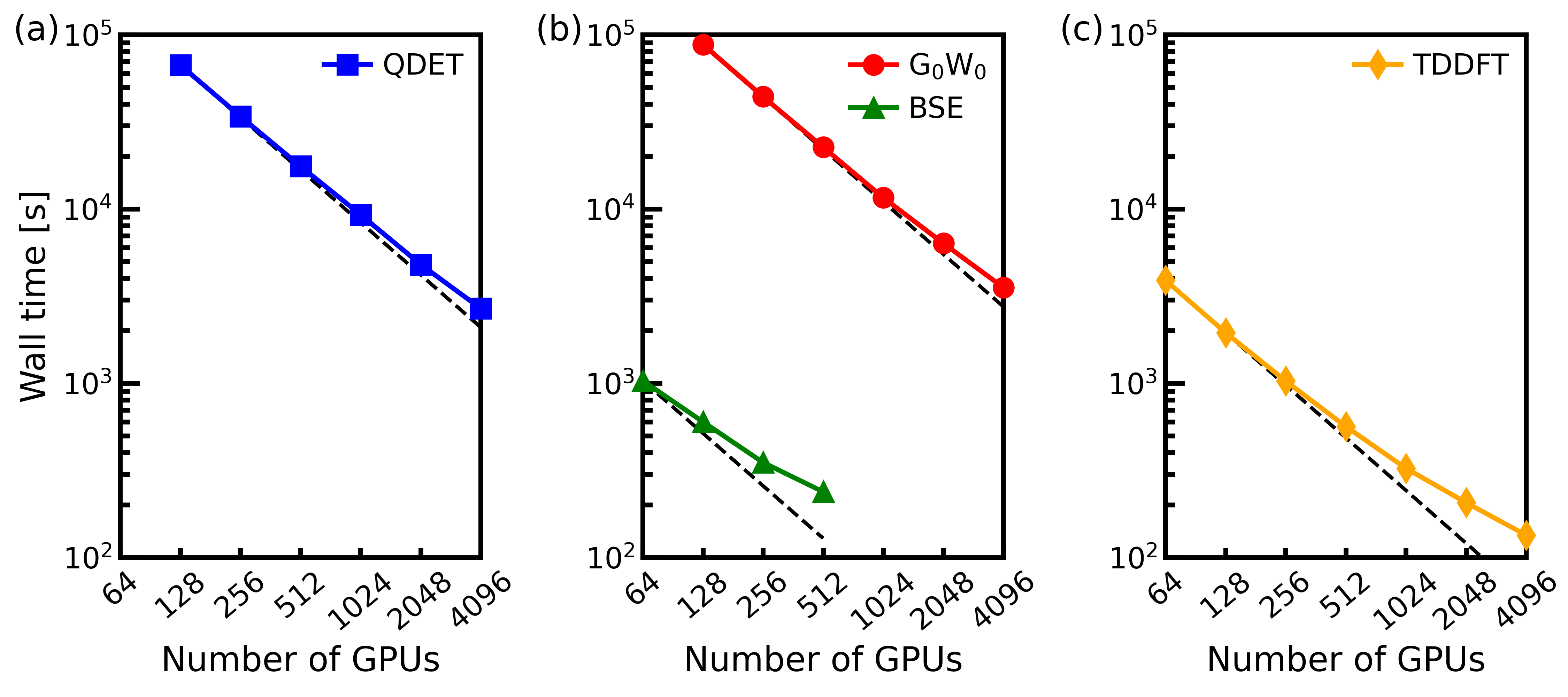}
\caption{Strong scaling of the WEST code benchmarked on the GPU partition of the Perlmutter supercomputer. Wall time is shown as a function of the number of GPUs for QDET (blue squares), G$_0$W$_0$ (red circles), BSE (green triangles), and TDDFT (orange diamonds) calculations of the nitrogen-vacancy center in a diamond $5 \times 5 \times 5$ supercell (999 atoms). Black dashed lines indicate ideal scaling. The QDET, G$_0$W$_0$, and BSE calculations used the PBE functional. The TDDFT calculation used the dielectric-dependent hybrid functional and computed excited-state analytical forces.}
\label{fig:parallel}
\end{figure*}

All capabilities of WEST described in section~\ref{sec:features}, including GW, QDET, BSE, TDDFT, and postprocessing functionalities, support GPU acceleration. Figure~\ref{fig:parallel} illustrates the excellent parallel efficiency of WEST on GPUs for representative QDET, G$_0$W$_0$, BSE, and TDDFT calculations of the negatively charged nitrogen-vacancy center (NV$^-$) in a diamond $5 \times 5 \times 5$ supercell with 999 atoms. The benchmarks were performed on the GPU partition of the Perlmutter supercomputer, where each compute node is equipped with four NVIDIA A100 GPUs. A plane-wave energy cutoff of 60 Ry was used in all calculations. The QDET, G$_0$W$_0$, and BSE calculations employed the PBE functional, whereas the TDDFT calculation employed the DDH functional and computed excited-state analytical forces. In all cases, WEST exhibits near-ideal scaling from 64 to 4,096 GPUs, highlighting the effectiveness of the hierarchical parallelization strategy combined with GPU acceleration. Minor deviations from ideal scaling at large GPU counts are attributed to input/output (I/O) and inter-node communication overhead. Overall, the performance benchmarks demonstrate the capability of WEST to efficiently exploit GPU-based supercomputers for large-scale simulations.

\subsection{Software development}

WEST is written in Fortran (90+) with a modular and extensible design for ease of development and maintenance. The code utilizes standard HPC libraries, such as MPI, BLAS, LAPACK, and FFT, as shown in the green inset of figure~\ref{fig:workflow}. Core numerical algorithms are implemented in a unified manner and shared across multiple functionalities. For example, the diagonalization of the symmetrized density-density response function $\bar{\chi}$ (equation~\ref{eq:pdep}) and the BSE/TDDFT Liouville superoperator (equation~\ref{eq:bse_tddft}) relies on a common implementation of the Davidson iterative diagonalization algorithm. This unified design ensures consistency across different methods while reducing code duplication.

WEST is developed as an open-source project under the GNU Lesser General Public License (LGPL), and is publicly available on GitHub~\cite{west_github}, where it benefits from community contributions through issue tracking and pull requests. Continuous integration is implemented using GitHub Actions, with automated testing pipelines that validate core functionalities and parallelization schemes for any changes to the code base. This infrastructure ensures code reliability, reproducibility, and rapid detection of regressions. Comprehensive documentation and tutorials~\cite{west_website} are provided to support both developers and users, promoting the adoption of the code within the broader computational materials science community.

\subsection{Interoperability}
\label{sec:interop}

WEST adopts standard, structured data formats to facilitate interoperability and automation. Both JSON and YAML formats are supported for input, providing human-readable and machine-friendly interfaces that are well-suited for scripting and workflow integration. In addition to the human-readable program summary printed to the terminal, computational results are also stored in JSON format, enabling seamless post-processing and data exchange. Parsing and writing of JSON and YAML files take advantage of the built-in functionality of Python and of a Fortran-Python interface layer.

WEST is interfaced with a growing ecosystem of external tools and libraries, shown as orange insets in figure~\ref{fig:workflow}. Notable integrations include the following:
\begin{itemize}
\item PySCF, for the diagonalization of QDET effective Hamiltonians on classical computers using FCI or SCI~\cite{pyscf_sun_2020}.

\item Qiskit, for the diagonalization of QDET effective Hamiltonians on quantum computing platforms. WEST is used to construct the QDET effective Hamiltonian, which is subsequently mapped onto a set of Pauli operators and diagonalized on a quantum computer using Qiskit with algorithms including the variational quantum eigensolver and quantum subspace expansion~\cite{quantum_huang_2022,quantum_vorwerk_2022,quantum_huang_2023}.

\item PyRET, for the simulation of non-resonant energy transfer processes between many-body states obtained from DFT or QDET calculations~\cite{ret_chattaraj_2024,ret_chattaraj_2025}.

\item PyPL, for the calculation of vibrationally resolved optical spectra based on Huang-Rhys theory and the generating function method~\cite{photoluminescence_jin_2021,tddft_jin_2022}. WEST is used to compute excited-state forces within BSE/TDDFT, enabling geometry relaxation and phonon calculations in the excited states.

\item PyFRSOC, for the evaluation of SOC matrix elements between many-body states obtained from QDET calculations, and the calculation of the SOC contribution to zero-field splitting matrix elements~\cite{isc_jin_2025,pyfrsoc_jin_2026}.

\item PyEPFD, for the computation of vibronic coupling. WEST is used to compute excited-state forces within BSE/TDDFT and excited-state phonon modes. PyEPFD then uses stochastic sampling techniques to evaluate vibronic effects. Excitation energies for stochastic configurations are averaged to obtain observables including the effect of vibronic coupling~\cite{vibronic_kundu_2024}.
\end{itemize}

\section{Representative applications of the WEST code}
\label{sec:applications}

In this section, we illustrate the key capabilities of WEST through representative applications, ranging from detailed studies of prototypical quantum defects to large-scale simulations and materials discovery. These examples highlight the accuracy, scalability, and versatility of the methodologies implemented in WEST. All DFT calculations were carried out using Quantum ESPRESSO.

\subsection{Access to multiple excited-state properties}

The NV$^-$ center in diamond, consisting of a substitutional nitrogen atom adjacent to a vacancy, is a paradigmatic solid-state quantum defect, with spin states that can be initialized and read out optically. Its localized defect states within the band gap enable an optical cycle involving a triplet ground state $^3A_2$, a triplet excited state $^3E$, and intermediate singlet states $^1E$ and $^1A_1$ that act as shelving states. Accurate characterization of vertical and adiabatic excitation energies and transition rates between these states is thus essential for understanding and optimizing the performance of the NV$^-$ center in quantum technologies. Here, we demonstrate how all methodologies implemented in WEST, G$_0$W$_0$, QDET, BSE, and TDDFT, can be applied coherently to computationally characterize the NV$^-$ center across multiple levels of theory and a broad range of physical observables and processes.

\subsubsection{Defect levels}

\begin{figure}[ht!]
\includegraphics[width=0.48\textwidth]{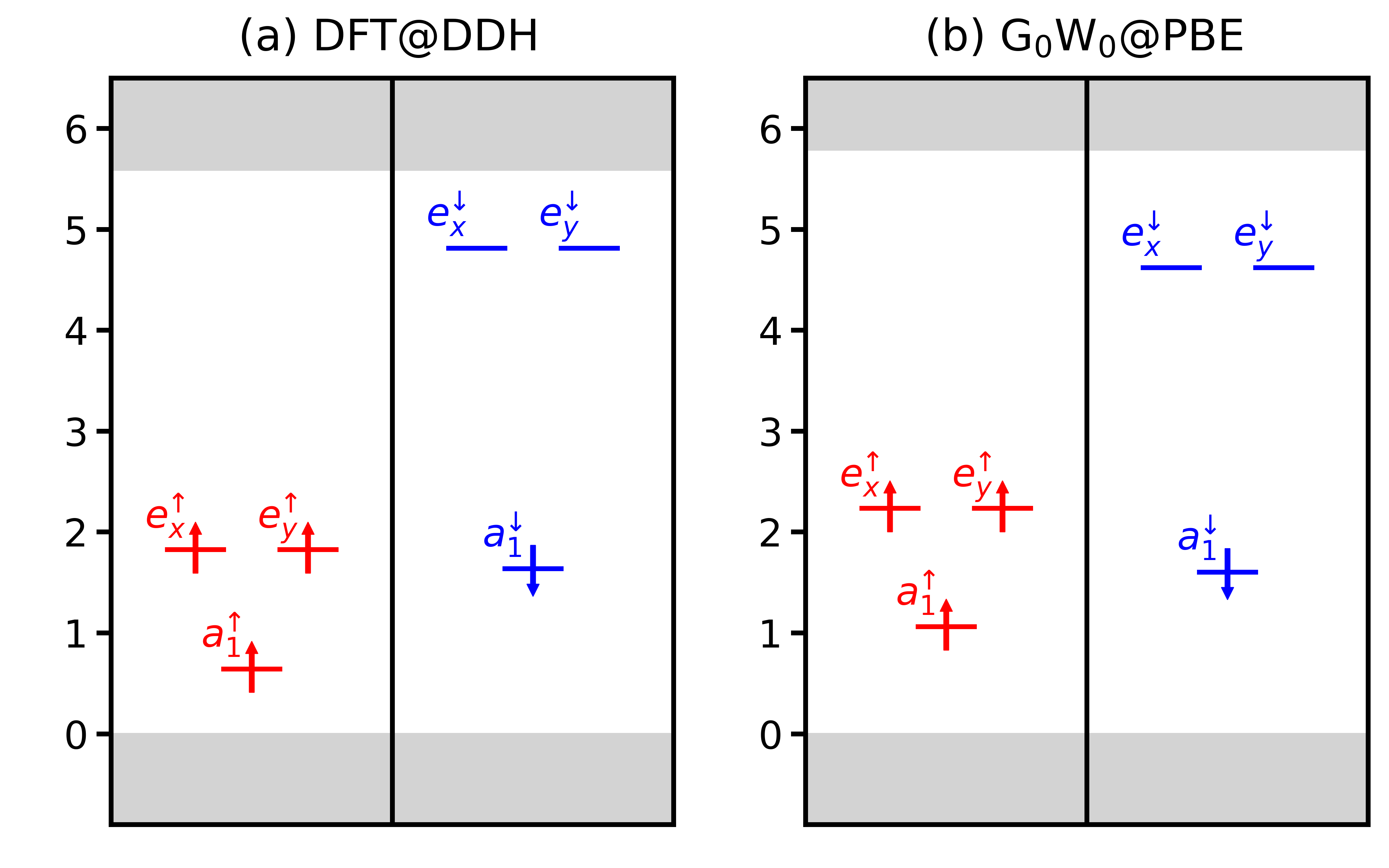}
\caption{Defect energy levels of the NV$^-$ center in a $4 \times 4 \times 4$ diamond supercell, computed using (a) DFT with the DDH functional, and (b) G$_0$W$_0$ with the PBE functional. Spin-up and spin-down energy levels are shown in red and blue, respectively, while the gray regions indicate the valence and conduction bands of diamond.}
\label{fig:nv_diamond_levels}
\end{figure}

We begin by examining the single-particle electronic structure of the NV$^-$ center. Figure~\ref{fig:nv_diamond_levels} compares band edges and defect energy levels obtained using hybrid DFT with the DDH functional (with 18\% exact exchange) and QP energies computed using G$_0$W$_0$ with the PBE functional. Both methods predict a band gap (5.59 eV for DFT@DDH and 5.79 eV for G$_0$W$_0$@PBE), in close agreement with the experimental value of 5.47 eV~\cite{diamond_clark_1964}, or 5.8 eV when zero-point renormalization (ZPR) is included~\cite{zpr_engel_2022,zpr_ponce_2025}. For the defect levels within the band gap, we find an overall consistency between DFT@DDH and G$_0$W$_0$@PBE results, with minor differences. In the spin-up channel, the $a_1^{\uparrow}$ and $e^{\uparrow}$ defect levels computed with DFT@DDH lie closer to the valence band maximum (VBM) than in the G$_0$W$_0$@PBE case. Conversely, in the spin-down channel, G$_0$W$_0$@PBE predicts the $a_1^{\downarrow}$ and $e^{\downarrow}$ levels to be closer to the VBM compared to DFT@DDH. The energy separation between the $e^{\downarrow}$ and $a_1^{\downarrow}$ levels is 3.18 eV at the DFT@DDH level and 3.02 eV at the G$_0$W$_0$@PBE level.

\subsubsection{Vertical excitation energies}

\begin{figure}[ht!]
\includegraphics[width=0.48\textwidth]{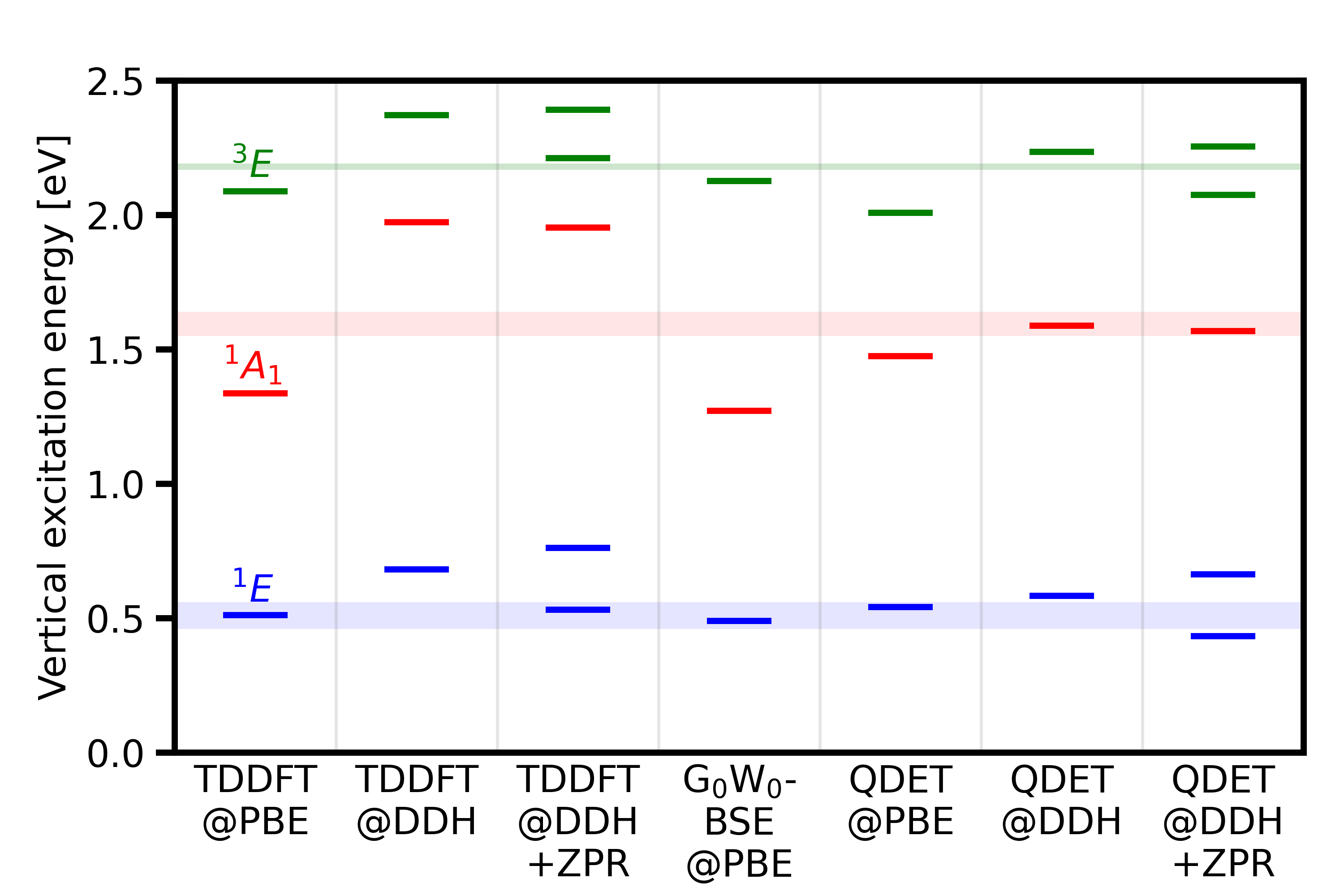}
\caption{Vertical excitation energies (VEEs) of the NV$^-$ center in a $4 \times 4 \times 4$ diamond supercell, computed using TDDFT with PBE and DDH functionals, G$_0$W$_0$-BSE with PBE, and QDET with PBE and DDH. TDDFT@DDH and QDET@DDH results including zero-point renormalization (ZPR)~\cite{vibronic_kundu_2024} are also shown. Shaded areas indicate experimentally inferred VEEs~\cite{nv_davies_1976,nv_rogers_2008,nv_goldman_2015a,nv_goldman_2015b}.}
\label{fig:nv_diamond_methods}
\end{figure}

Starting from a single-particle picture, we computed the VEEs of the $^1E$, $^1A_1$, and $^3E$ excited states using TDDFT, BSE, and QDET, with the singlet states obtained with the spin-flip TDDFT and BSE implementations. The computed $\langle S^2 \rangle$ values yield the correct spin multiplicity for each state. In figure~\ref{fig:nv_diamond_methods}, we compare VEEs computed with different methods and XC functionals. Overall, all methods yield qualitatively consistent results. Notably, VEEs obtained with QDET exhibit a weaker dependence on the choice of XC functional than those obtained with TDDFT. When comparing computed VEEs with experimental data, it is important to account for effects of nuclear quantum motion, which is known to lift the degeneracy of the $^1E$ and $^3E$ states and to shift the energies of the $^1E$, $^1A_1$, and $^3E$ states by as much as 0.16 eV~\cite{vibronic_kundu_2024}. After including ZPR corrections, VEEs obtained with TDDFT@DDH and QDET@DDH are in good agreement with experiment. The ZPR is evaluated by performing TDDFT, BSE, or QDET calculations on tens to hundreds of stochastically sampled nuclear configurations, a computationally demanding task made feasible by the efficiency of WEST. The remaining discrepancies between theory and experiment may originate from several approximations. For TDDFT and BSE, these include the adoption of the TDA, the neglect of dynamical screening effects~\cite{bse_rocca_2010}, and the restriction to single excitations. For QDET, errors may arise from the use of a frequency-independent effective Hamiltonian and the absence of a self-consistent update of the active space and environment during the embedding procedure.

We emphasize that the comparison of methods presented in figure~\ref{fig:nv_diamond_methods} is carried out using identical geometries, basis sets, pseudopotentials, and other numerical parameters. In contrast, comparisons reported in the literature are often based on results obtained with different codes and computational setups, where variations in numerical approximations and implementations can obscure the intrinsic differences between the underlying levels of theory.

\begin{figure*}[ht!]
\includegraphics[width=0.72\textwidth]{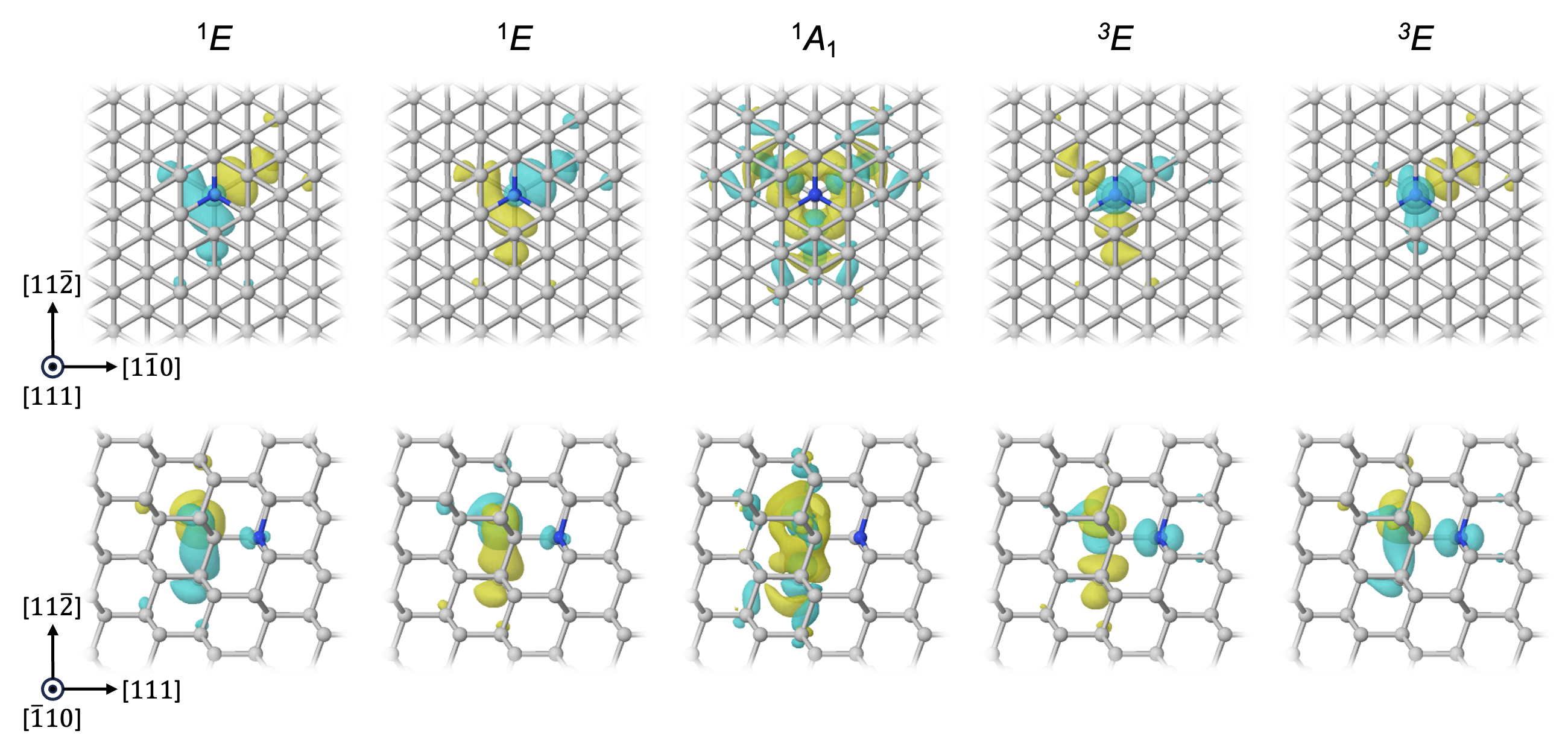}
\caption{Unrelaxed differential density of the $^1E$, $^1A_1$, and $^3E$ excited states of the NV$^-$ center in a $4 \times 4 \times 4$ diamond supercell, computed using TDDFT@DDH. The top and bottom panels correspond to views along the [111] and [$\bar{1}$10] directions, respectively. Carbon atoms are shown in gray, whereas nitrogen atoms are shown in blue. Blue and yellow regions represent electron depletion and accumulation, respectively.}
\label{fig:nv_diamond_diff_rho}
\end{figure*}

Figure~\ref{fig:nv_diamond_diff_rho} visualizes the unrelaxed differential density, defined in equation~\ref{eq:diff_rho} and computed using TDDFT@DDH, for the $^1E$, $^1A_1$, and $^3E$ excited states. It is known that the square modulus of the wave function of the $a_1$ defect level extends both on the substitutional nitrogen and the three carbon atoms adjacent to the vacancy, whereas those of the $e$ defect levels are primarily localized on the three carbon atoms surrounding the vacancy. The isosurfaces plotted in figure~\ref{fig:nv_diamond_diff_rho} reveal that the $^1E$ and $^1A_1$ states are dominated by transitions within the $e$ manifold, while the $^3E$ states involve transitions from the $a_1$ level into the $e$ levels. This interpretation is consistent with an analysis based on decomposing the excited states into electron-hole pairs following equation~\ref{eq:decomp}.

\subsubsection{Excited-state geometries}

Excited-state structural relaxation plays a central role in the optical cycle of the NV$^-$ center, including in determining the zero-phonon line (ZPL) and photoluminescence (PL). Upon excitation from the $^3A_2$ ground state to the $^3E$ excited state, the defect undergoes a Jahn–Teller distortion that lowers the $C_{3v}$ symmetry. We relaxed excited-state geometries using both $\Delta$SCF and TDDFT. For the triplet $^3E$ state, both approaches reproduce the ZPL energy in fair agreement with the experimental value of 1.945 eV~\cite{nv_davies_1976}. Specifically, $\Delta$SCF using the PBE and DDH functionals yields 1.71 and 2.21 eV, respectively, while TDDFT using the PBE and DDH functionals yields 1.89 and 2.11 eV, respectively~\cite{tddft_jin_2023}. It should be noted, however, that such an agreement between $\Delta$SCF and TDDFT is not generally expected across all systems.

\begin{figure}[ht!]
\includegraphics[width=0.48\textwidth]{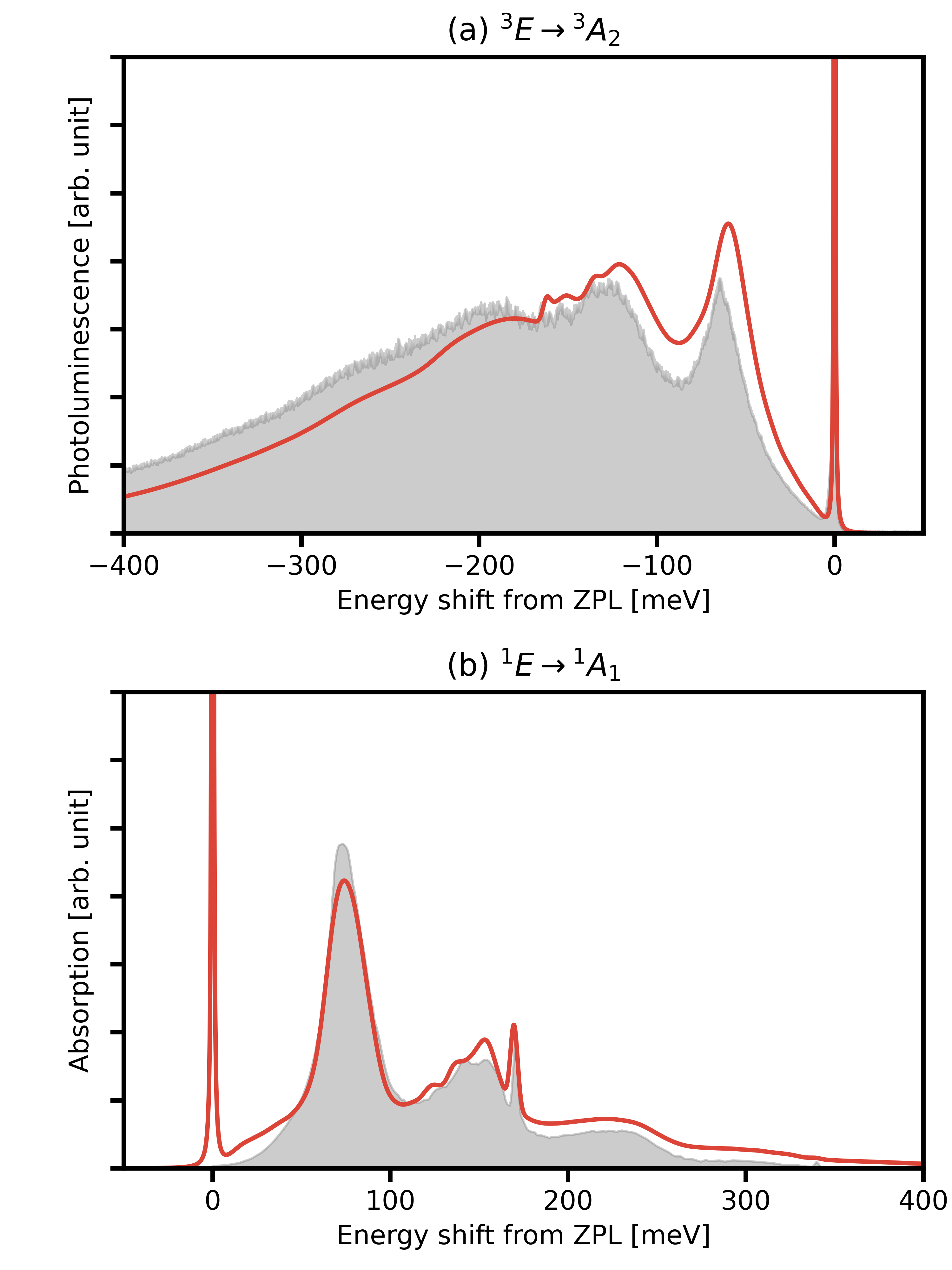}
\caption{(a) Photoluminescence (PL) line shape of the $^3E \rightarrow {}^3A_2$ transition and (b) absorption line shape of the $^1E \rightarrow {}^1A_1$ transition of the NV$^-$ center in diamond. Red lines represent computational results, while gray areas represent experimental data~\cite{nv_alkauskas_2014,nv_kehayias_2013}. Note that the ZPL of the $^1E \rightarrow {}^1A_1$ transition exists in experiment but is not depicted in the experimental spectrum. Computational results are extrapolated to the dilute limit. Reproduced from reference~\onlinecite{tddft_jin_2022}, available under a CC BY 4.0 license. Copyright 2022 Jin et al.}
\label{fig:nv_diamond_spectra}
\end{figure}

Using the TDDFT@DDH optimized excited-state geometry together with phonons computed at the DFT@PBE level, we further evaluated the PL spectrum of the $^3E \rightarrow {}^3A_2$ transition, shown in figure~\ref{fig:nv_diamond_spectra}(a). The resulting phonon sideband reproduces the main peaks and features of the experimental line shape, indicating that the excited-state geometry relaxation and e-ph interactions are well captured at this level of theory. Unlike the $^3E$ state that can be described with a single Slater determinant, the $^1E$ and $^1A_1$ singlet states cannot be treated within the standard $\Delta$SCF framework due to their multi-configurational character. We optimized their geometries using spin-flip TDDFT, and computed the absorption spectrum of the $^1E \rightarrow {}^1A_1$ transition, as shown in figure~\ref{fig:nv_diamond_spectra}(b). The computed line shape again reproduces the key features of the experimental spectrum, demonstrating the good performance of the TDDFT excited-state relaxation procedure for both spin-conserving and spin-flip excitations.

\subsubsection{Radiative and non-radiative decay rates}

By combining the various capabilities of the WEST code, we are able to fully characterize, from first principles, all transitions involved in the optical cycle of spin defects. Here, we present results for the visible and infrared fluorescence of the NV$^-$ center. The fluorescence is quantified by the lifetime
\begin{equation}
\label{eq:fluorescence}
\tau (T) = \frac{1}{\Gamma_{\mathrm{rad}} + \Gamma_{\mathrm{nonrad}} (T)} \,,
\end{equation}
which reflects the competition between radiative and non-radiative decays. The radiative decay rate $\Gamma_{\mathrm{rad}}$ is primarily determined by the transition dipole moment between $^3E$ and $^3A_2$, which can be computed by TDDFT or BSE using
\begin{equation}
\Gamma_{\mathrm{rad}} = \frac{n \vert E_{\mathrm{ZPL}} \vert ^3 \vert \mu \vert ^2} {3 \pi \epsilon_0 c^3 \hbar^4} \,,
\end{equation}
where $n$ is the refractive index of the material (2.4 for diamond), $E_{\mathrm{ZPL}}$ is the ZPL energy, $\mu$ is the transition dipole moment computed using equation~\ref{eq:tdm}, $\epsilon_0$ is the vacuum permittivity, $c$ is the speed of light, and $\hbar$ is the reduced Planck constant. Using $\mu$ computed with TDDFT@DDH and G$_0$W$_0$@PBE, we obtained $\Gamma_{\mathrm{rad}}$ values of 49.8 and 84.3 MHz, respectively. The G$_0$W$_0$@PBE result agrees well with the experimental value of 83 MHz~\cite{nv_doherty_2013,nv_goldman_2015a}, whereas TDDFT@DDH underestimates the experimental value by more than 30\%, since the contribution of the exact exchange is omitted when evaluating the commutator $[H_{\mathrm{KS}}, \mathbf{r}]$.

We consider two principal non-radiative decay channels, namely the spin-changing ISC induced by SOC and the spin-conserving IC induced by e-ph coupling. The temperature-dependent non-radiative decay rate, $\Gamma_{\mathrm{nonrad}}(T) = \Gamma_{\mathrm{ISC}} (T) + \Gamma_{\mathrm{IC}} (T)$, is the sum of the ISC and IC contributions. The ISC rate $\Gamma_{\mathrm{ISC}} (T)$ is governed by transitions between the triplet and singlet manifolds and depends sensitively on the zero-phonon energy gaps, the vibrational overlap functions, and the strength of the SOC matrix elements between the states. The IC rate $\Gamma_{\mathrm{IC}}(T)$ is governed by transitions within the same spin manifold and depends on the strength of the e-ph coupling, as well as on the zero-phonon energy gaps and the vibrational overlap functions. Owing to the complexity of the ISC and ISC processes, the quantitative prediction of non-radiative decay rates is considerably more challenging than that of radiative decay rates. Recently, we developed computational frameworks~\cite{isc_jin_2025,ic_villani_2026} that combine multiple functionalities of the WEST code and its interoperable software modules (section~\ref{sec:interop}) to compute ISC and IC rates from first principles. Specifically, SOC matrix elements are evaluated between many-body states obtained from QDET, while e-ph matrix elements are computed using the NACs between many-body states obtained from TDDFT. The implementation of analytical NACs eliminates the bottleneck of finite-difference calculations of wave-function derivatives, enabling the inclusion of all phonon modes. The zero-phonon energy gaps are determined from excited-state geometries relaxed using TDDFT. The excited-state geometries are also used, together with phonons computed from DFT and TDDFT, to construct the vibrational overlap functions. The computed IC rates are substantially lower than the radiative and ISC contributions, indicating that the non-radiative decay between triplet states is dominated by ISC. For example, over the temperature range of 100 to 400 K, $\Gamma_{\mathrm{IC}} \lesssim 5$ MHz, whereas $\Gamma_{\mathrm{ISC}}$ ranges from 54 to 61 MHz.

\begin{figure}[ht!]
\includegraphics[width=0.48\textwidth]{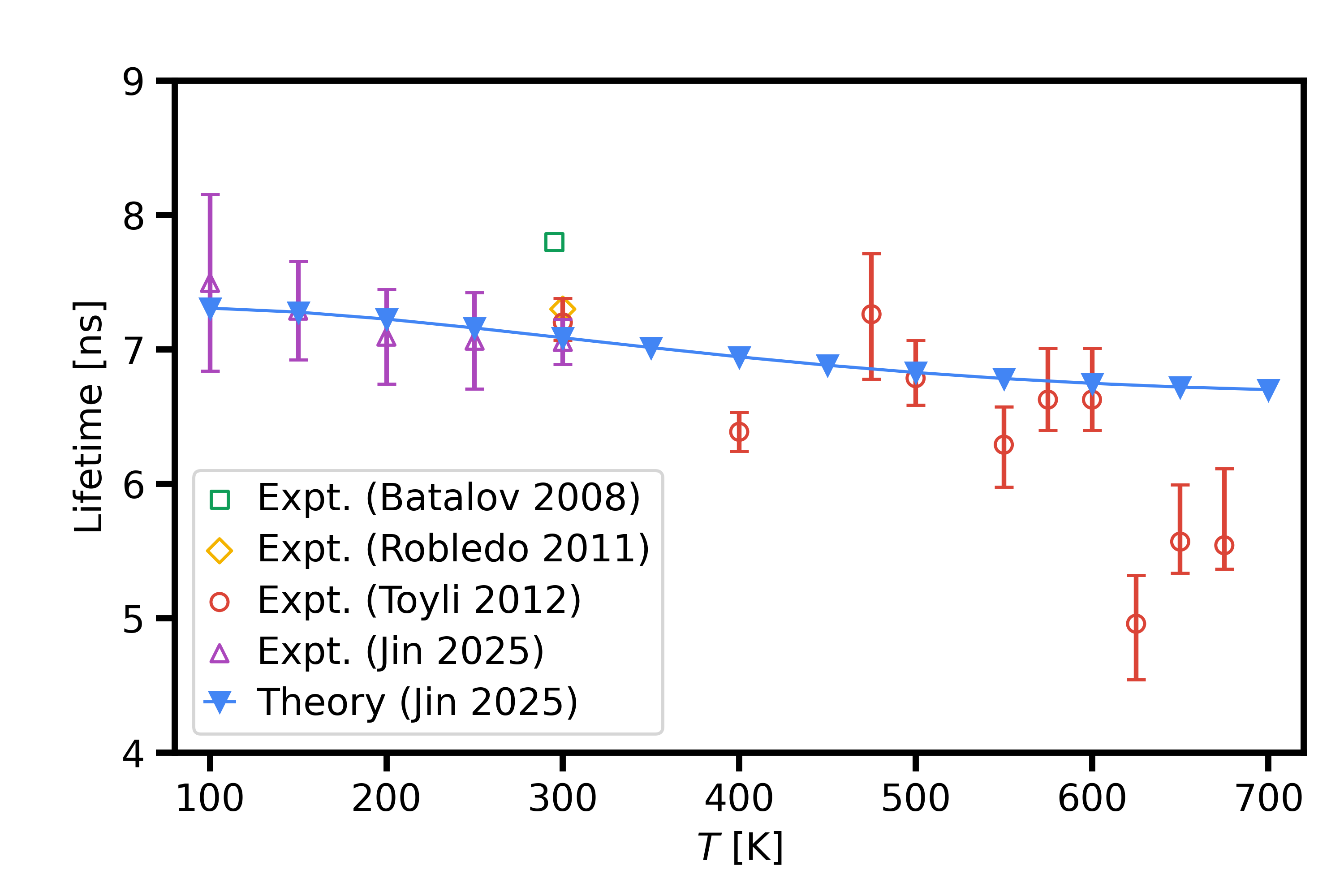}
\caption{Computed fluorescence lifetimes (filled blue triangles) of the $m_S = \pm 1$ sublevels of the $^3E$ excited state of the NV$^-$ center in diamond as a function of temperature ($T$), compared with experimental values (open symbols) from Batalov et al.~\cite{isc_batalov_2008}, Robledo et al.~\cite{isc_robledo_2011}, Toyli et al.~\cite{isc_toyli_2012}, and Jin et al.~\cite{isc_jin_2025} Reproduced with permission from reference~\onlinecite{isc_jin_2025}. Copyright 2025, American Physical Society.}
\label{fig:nv_diamond_isc}
\end{figure}

Using equation~\ref{eq:fluorescence}, we obtained the overall fluorescence rate and the corresponding lifetime, and compared the latter with experimental measurements~\cite{isc_batalov_2008,isc_robledo_2011,isc_toyli_2012,isc_jin_2025}, as shown in figure~\ref{fig:nv_diamond_isc}. The results demonstrate that the temperature dependence of the fluorescence lifetime is well reproduced up to room temperature. We also computed, for the first time, the lifetime of the $^1A_1$ state, whose decay to the lower singlet state $^1E$ is instead dominated by the non-radiative IC processes. We obtained a value of 117.4 ps~\cite{ic_villani_2026}, in excellent agreement with the experimentally inferred lifetime of $\sim$100 ps at 78 K, as measured from femtosecond transient infrared absorption spectroscopy measurements~\cite{nv_ulbricht_2018}.

Taken together, these results illustrate the breadth of excited-state properties and processes that can be simulated with WEST. Crucially, these quantities are treated consistently within a unified framework.

\subsection{Scalability to large systems}
\label{sec:large}

The accurate theoretical description of realistic material environments often requires supercells containing more than a thousand atoms in order to capture finite-size effects, long-range interactions, and structural heterogeneities. The scalable implementations of TDDFT, BSE, and QDET in WEST enable calculations for such large systems representing experimentally relevant environments. In the following, we illustrate these capabilities through two examples, including a rigorous study of finite-size effects on VEEs and a thorough investigation of spin defects in the vicinity of mesoscopic defects.

\subsubsection{Finite-size effects on vertical excitation energies}

\begin{figure}[ht!]
\includegraphics[width=0.4\textwidth]{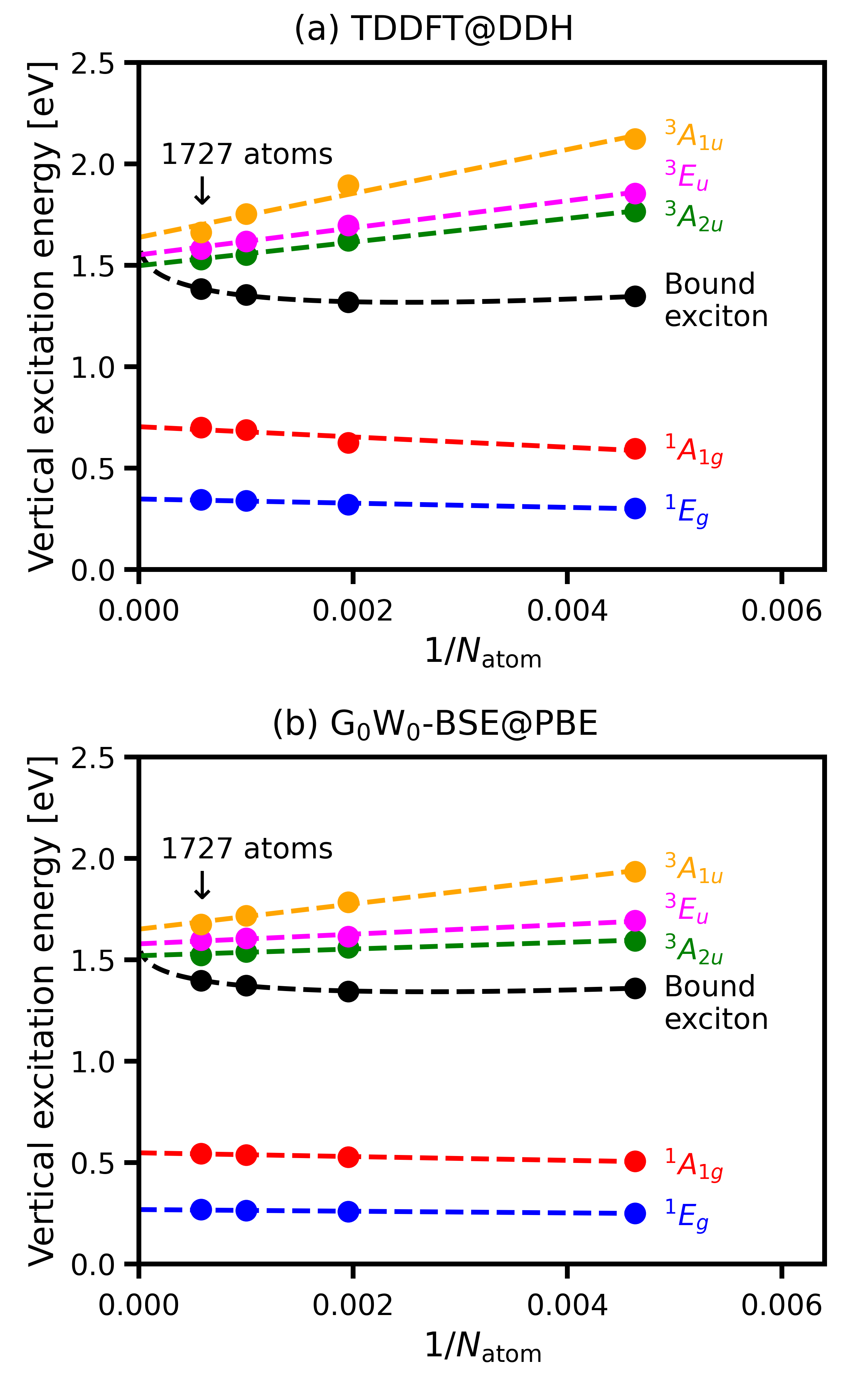}
\caption{Vertical excitation energies (VEEs) of the $^1E_g$, $^1A_{1g}$, $^3A_{2u}$, $^3E_u$, and $^3A_{1u}$ excited states and the bound exciton state of the SiV$^0$ center in diamond, computed using (a) TDDFT@DDH and (b) G$_0$W$_0$-BSE@PBE. The VEEs are shown as a function of the number of atoms ($N_{\mathrm{atom}}$) in the supercell. The dashed lines indicate extrapolations as a function of $1/N_{\mathrm{atom}}$, using the extrapolation defined in equation~\ref{eq:bound_exciton} for the bound exciton, and a linear extrapolation for the other excited states. Adapted with permission from reference~\onlinecite{bse_yu_2024}. Copyright 2024, American Chemical Society.}
\label{fig:siv_diamond}
\end{figure}

The neutral silicon vacancy center (SiV$^0$) in diamond, consisting of a substitutional silicon adjacent to a vacancy, exhibits long spin coherence time along with a near-infrared fluorescence signal and has emerged as a promising platform for quantum communication applications~\cite{siv_rose_2018,siv_green_2019}. This defect possesses $D_{3d}$ symmetry, with the $e_g$ type defect levels localized in the band gap of diamond and the $e_u$ type defect levels resonant with the valence bands. We performed TDDFT@DDH and G$_0$W$_0$-BSE@PBE calculations using supercells with different numbers of atoms to assess finite-size effects on the computed VEEs. The results are shown in figure~\ref{fig:siv_diamond}. The VEEs of the triplet states exhibit an approximately linear dependence on $1/N_{\mathrm{atom}}$, arising from dipole-dipole interactions between localized excitons in periodic images. In contrast, the VEEs of the singlet states show only weak dependence on supercell size, as their many-body wave functions are primarily composed of transitions between defect levels within the band gap of diamond.

Bound excitons (BEs) have been observed experimentally with excitation energies comparable to the ZPL of the triplet excited states~\cite{siv_zhang_2020}. The VEE of the BE, computed as the transition from the VBM to the $e_{gx}^{\downarrow}$ and $e_{gy}^{\downarrow}$ defect levels, exhibits a non-linear dependence on the supercell size, which can be attributed to the combined effects of the dipole-dipole interaction between excitons in nearby periodic images and the electron-hole interaction within the same exciton. We fitted the VEE of the BE using the function
\begin{equation}
\label{eq:bound_exciton}
E_{\mathrm{BE}}(L) = E_{\mathrm{BE}}(L = \infty) + \frac{A}{L} \exp \left(-\frac{L}{D}\right) - \frac{B}{L^3} \,,
\end{equation}
where $L \propto N_{\mathrm{atom}}^{1/3}$ is the length of the cubic supercell, $A$ and $B$ are fitting parameters determined by the strength of the electron-hole interaction and the dipole-dipole interaction, respectively. The screening parameter $D$ takes into account the fact that the electron-hole interaction no longer depends on $L$ when $L$ is much larger than the radius of the bound exciton. Using $D \in [10, 40]$ \AA~\cite{siv_zhang_2020,tddft_jin_2023}, we obtained $E_{\mathrm{BE}}(L = \infty) \in [1.33, 1.50]$ eV, in close agreement with the experimental value of 1.39 eV~\cite{siv_zhang_2020}. This example shows that the affordable computational cost of TDDFT and G$_0$W$_0$-BSE in WEST has enabled a systematic assessment of finite-size effects.

\subsubsection{Complex supercells representing heterogeneous environments}

\begin{figure}[ht!]
\includegraphics[width=0.48\textwidth]{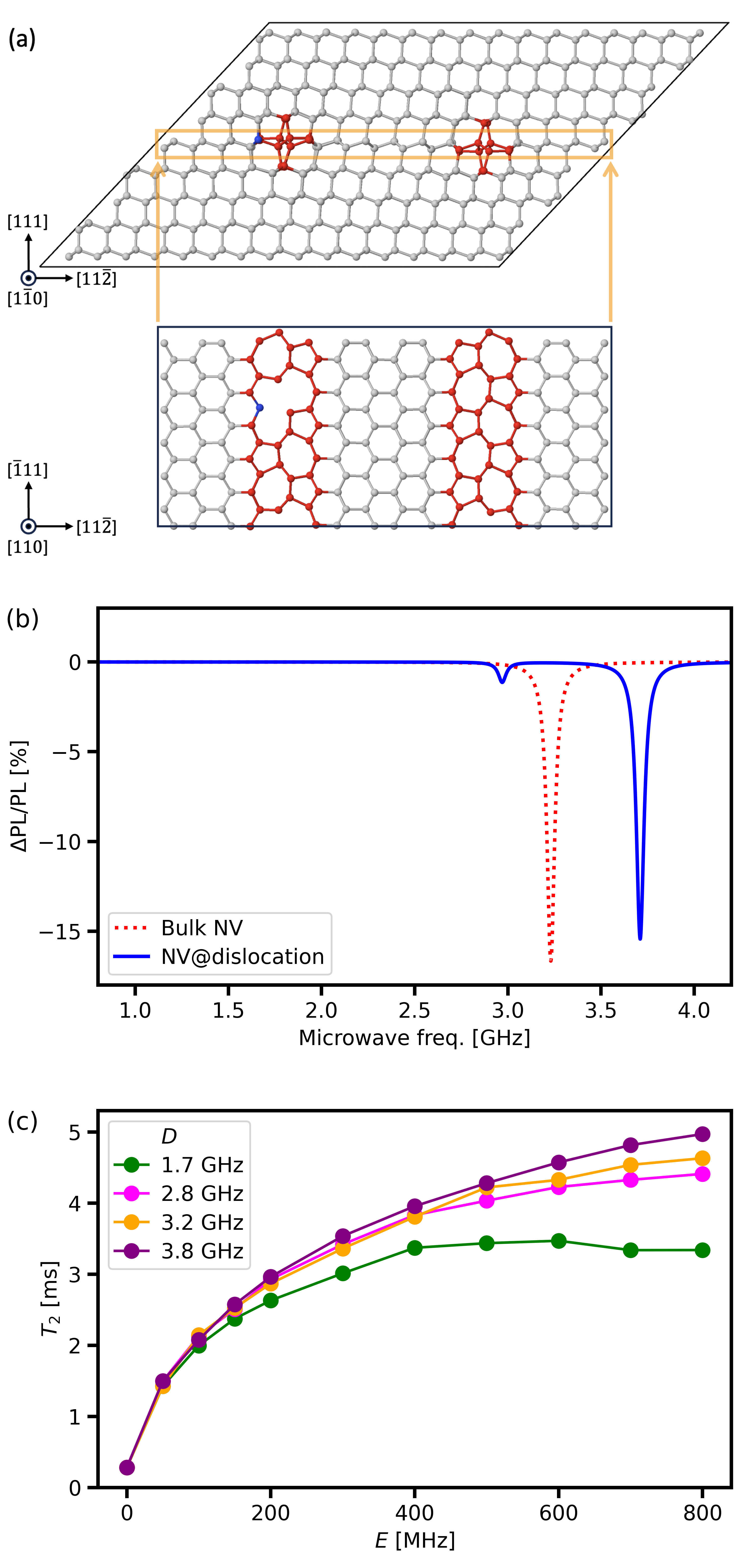}
\caption{(a) Diamond supercell including two dislocations oriented in opposite directions and one NV$^-$ center. Carbon atoms at the dislocation cores are shown in red, while the rest of the carbon atoms are shown in gray. Nitrogen atoms are shown in blue. (b) Simulated continuous-wave optically detected magnetic resonance (ODMR) spectra of NV$^-$ in bulk diamond and a representative NV$^-$ at the dislocation core. (c) Predicted coherence times ($T_2$) of NV$^-$ centers at dislocations. $T_2$ is computed at zero magnetic field as a function of the transverse component ($E$) of the zero-field splitting (ZFS) tensor, considering different values of the axial component ($D$) of the ZFS. The point at $E = 0$ corresponds to NV$^-$ in bulk diamond, whereas points with $E > 0$ correspond to NV$^-$ centers at the dislocation core. Adapted with permission from reference~\onlinecite{dislocation_zhang_2026}. Copyright 2026, Springer Nature.}
\label{fig:nv_diamond_dislocation}
\end{figure}

In the solid state, controlling and scaling defect-based qubits into coherent, interconnected arrays remains a major challenge. In principle, arrays of spin defects may be engineered by exploiting the strain fields induced by dislocations in crystals. To explore this possibility, we investigated the feasibility of using NV centers close to dislocations in diamond as spin qubits~\cite{dislocation_zhang_2026}. Due to periodic boundary conditions employed in our simulations, we used a supercell of diamond with 1,727 atoms, containing two oppositely oriented glide dislocations and one NV center positioned near the core of one dislocation, as illustrated in figure~\ref{fig:nv_diamond_dislocation}(a).

Our formation energy calculations based on hybrid DFT revealed that NV centers are energetically attracted to dislocation cores, suggesting that dislocations can indeed serve as scaffolds for assembling NV centers into one-dimensional arrays. We found that multiple NV configurations near the dislocation core can be stabilized in the desired negatively charged triplet spin state. We further computed the electronic, optical, and coherence properties of representative NV configurations near the dislocation core, including the VEE, ZPL, ISC rates, zero-field-splitting (ZFS) tensor, and spin coherence time. The ISC rates, obtained by combining DFT, TDDFT, and QDET calculations, were subsequently used in state population simulations to model the optical cycle of the qubit initialization and readout, which were predicted to remain viable for several configurations. One example is shown in figure~\ref{fig:nv_diamond_dislocation}(b), where the NV center in bulk diamond and a representative NV configuration at the dislocation core exhibit comparable continuous-wave optically detected magnetic resonance (ODMR) contrast. Certain NV configurations can even outperform NV centers in bulk diamond, exhibiting larger Debye-Waller factors and longer coherence times, as shown in figure~\ref{fig:nv_diamond_dislocation}(c).

We emphasize that the study of the optical cycle of spin defects near dislocations was enabled by accurate, scalable, and interoperable simulation capabilities of the WEST code, summarized in figures~\ref{fig:workflow} and \ref{fig:parallel}. The study reported in reference~\onlinecite{dislocation_zhang_2026} represents the first study of the interaction between a point defect and a mesoscopic defect at a level of theory beyond DFT, using TDDFT and QDET. A similar computational study of ISC rates and optical cycles has also been performed for NV$^-$ centers in diamond under pressure, where simulations helped identify strategies to optimize the performance of NV-based high-pressure sensors through the control of local stress environments~\cite{isc_huang_2026}. The computational framework is general and can be readily extended to other defect platforms, such as divacancies in proximity to dislocations and interfaces in silicon carbide.

\subsection{Applications across material classes}

In this section, we highlight applications to broad classes of systems of the various methods implemented in WEST. These include spin defects in wide-band-gap insulators, low-dimensional materials, and molecular systems, metal-halide perovskites, water, and ice. We describe different properties, including band gaps, excitation energies, absorption and PL spectra, and self-trapped excitons (STEs).

\subsubsection{Spin qubits in emerging platforms}

\begin{figure}[ht!]
\includegraphics[width=0.48\textwidth]{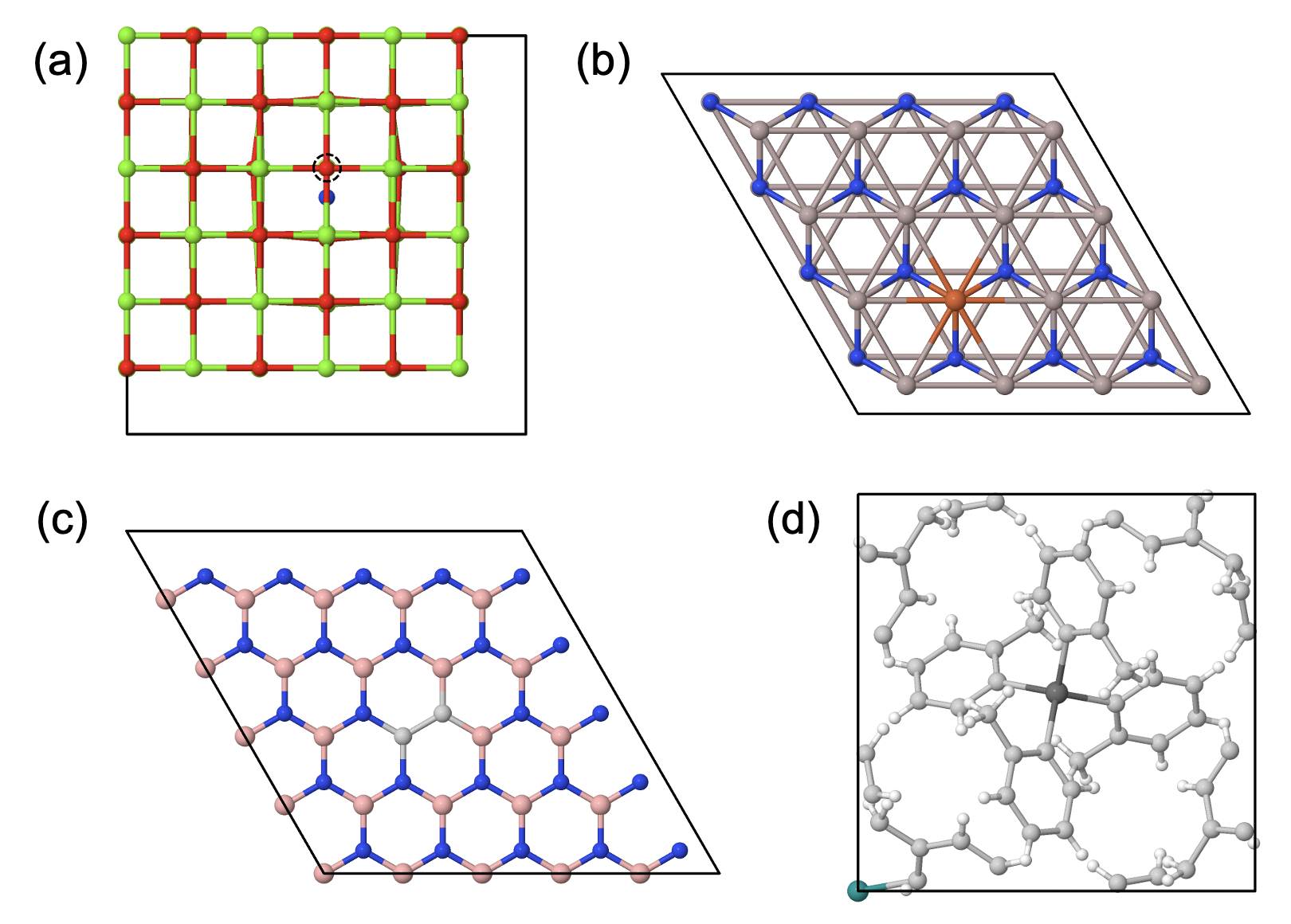}
\caption{(a) Negatively charged nitrogen-interstitial magnesium-vacancy complex ($\mathrm{[N_iV_{Mg}]^-}$) in magnesium oxide. (b) Iron substituting aluminum (Fe$_{\mathrm{Al}}$) in wurtzite aluminum nitride. (c) Carbon dimer (C$_{\mathrm{B}}$C$_{\mathrm{N}}$) in monolayer hexagonal boron nitride. (d) Cr(\textit{o}-tolyl)$_4$ molecular qubit. Hydrogen, boron, carbon, nitrogen, oxygen, magnesium, aluminum, chromium, iron, and tin atoms are shown in white, pink, gray, blue, red, green, taupe, dark gray, orange, and teal, respectively.}
\label{fig:spin_defects}
\end{figure}

While diamond and silicon carbide remain leading hosts for solid-state spin qubits, recently increased attention has focused on alternative platforms, including wide-band-gap oxides and nitrides, two-dimensional (2D) materials, and molecular qubits, as illustrated in figure~\ref{fig:spin_defects}.

Oxides have been predicted to host spin defects with exceptionally long coherence times owing to their low abundance of spinful nuclei~\cite{coherence_kanai_2022}. However, experimentally identified spin defects in oxides remain scarce. Accurate theoretical characterization of the ground- and excited-state properties of spin defects is crucial to enable their experimental synthesis and identification. WEST has been used to investigate promising defects in oxides identified through high-throughput screening~\cite{adaq_davidsson_2021}. One example is the negatively charged nitrogen-interstitial magnesium-vacancy complex ($\mathrm{[N_iV_{Mg}]^-}$) in magnesium oxide (MgO)~\cite{mgo_somjit_2025}, for which QDET calculations revealed a many-body level structure closely analogous to that of the NV$^-$ center in diamond, confirming the presence of triplet ground and excited states together with singlet shelving states. These calculations constitute a critical step in understanding the feasibility of an optical cycle. Furthermore, hybrid TDDFT calculations of the relaxed triplet excited state enabled the analysis of vibronic coupling and PL spectra, providing insights into strategies for improving the optical performance of this defect. Computational studies using QDET and TDDFT have also been performed for $\mathrm{[X_{Ca}V_O]^-}$ defects in calcium oxide (CaO), which consist of a missing Ca-O pair and a substitutional dopant atom (antimony, bismuth, or iodine)~\cite{cao_davidsson_2024}, as well as for neutral F centers in MgO~\cite{tddft_jin_2023,mgo_verma_2023}.

In addition to oxides, several nitrides are emerging as interesting platforms for spin defects. We recently studied iron substituting aluminum (Fe$_{\mathrm{Al}}$) in wurtzite aluminum nitride, namely a transition metal impurity in an insulator, which has proven to be a challenging case for quantum embedding approaches~\cite{aln_muechler_2022,aln_kleiner_2025,aln_otis_2025}. We found that QDET accurately describes the strongly correlated high-spin ground state and low-lying excited states, while spin-flip TDDFT predicts PL spectra in good agreement with experiment, significantly improving upon DFT predictions~\cite{aln_otis_2025}.

We now turn to discuss different classes of materials, including 2D solids and molecular qubits. Spin defects in 2D materials such as hexagonal boron nitride (hBN) are attracting considerable interest because they combine the favorable properties of solid-state spin qubits with the unique advantages of atomically thin materials, including ultimate proximity to the target being sensed, and seamless integration into van der Waals heterostructures. As an example, we consider the carbon dimer (C$_{\mathrm{B}}$C$_{\mathrm{N}}$) in monolayer hBN, for which hybrid TDDFT calculations employing the SE-RSH functional yield a VEE of 4.85 eV~\cite{sersh_zhan_2026}, improving upon the value of 5.02 eV obtained using a conventional PBE0-like (with 41\% exact exchange) hybrid functional and agreeing better with the experimental value of 4.54 eV. Work is currently underway to evaluate the ZPR of the VEE, which is expected to lower the computed VEE and further improve agreement with experiment. The superior performance of SE-RSH arises from the use of a spatially dependent fraction of exact exchange and range-separation parameter, enabling an accurate description of the nonuniform dielectric screening in 2D materials. In contrast, the PBE0-like functional employs a spatially uniform fraction of exact exchange, whose value is typically determined empirically by fitting computed and experimental observables or by enforcing theoretical constraints such as the generalized Koopmans' condition~\cite{hbn_smart_2018}. The SE-RSH functional is free of empirical parameters and eliminates the need for any system-specific tuning. Furthermore, TDDFT calculations using SE-RSH have been shown to yield accurate VEEs for defects in bulk solids~\cite{sersh_zhan_2026}, demonstrating that the functional performs reliably for materials with different dielectric environments.

Molecular qubits, where quantum information is encoded in spin states of chemically synthesized molecules, offer a complementary approach to solid-state qubits. Their electronic properties can be tailored through the choice of ligands, metal centers, and molecular geometries, while enabling scalable synthesis and integration into crystals, surfaces, and metal-organic frameworks. We investigated the electronic properties of the molecular qubit Cr(\textit{o}-tolyl)$_4$, for which a ZPL energy of 1.194 eV was predicted by combining the VEE computed using QDET and the excited-state Franck-Condon shift computed using spin-flip TDDFT~\cite{qdet_chen_2025}. The computed ZPL value is in excellent agreement with the experimental value of 1.210 eV~\cite{molecular_bayliss_2020}.

\subsubsection{Self-trapped excitons in metal-halide perovskites}

Metal-halide perovskites are a versatile class of optoelectronic materials due to the remarkable tunability of their properties through chemical composition and dimensionality. A notable feature of many metal-halide perovskites is the formation of STEs, where an exciton becomes localized through a lattice distortion induced by the exciton itself. Radiative recombination of STEs typically gives rise to broadband emission with a large Stokes shift.

The hybrid TDDFT implementation in WEST provides a powerful framework for studying STEs by combining an explicit treatment of excitonic effects with excited-state geometry relaxations. This capability was employed to investigate STEs in Cs$_4$SnBr$_6$ and Cs$_2$AgInCl$_6$, promising candidates for efficient light-emitting materials, using TDDFT with the DDH functional~\cite{perovskite_jin_2024}. The calculations accurately reproduced optical gaps, STE emission energies, and broadband emission spectra in agreement with experiment. Importantly, the study showed that the properties of STEs cannot be understood solely from exciton-phonon coupling evaluated at the ground-state geometry. Instead, explicit relaxation of the excited-state potential-energy surface is essential to capture the large structural distortions associated with self-trapping. The use of analytical TDDFT excited-state forces also overcame convergence challenges encountered in $\Delta$SCF calculations.

\begin{figure}[ht!]
\includegraphics[width=0.48\textwidth]{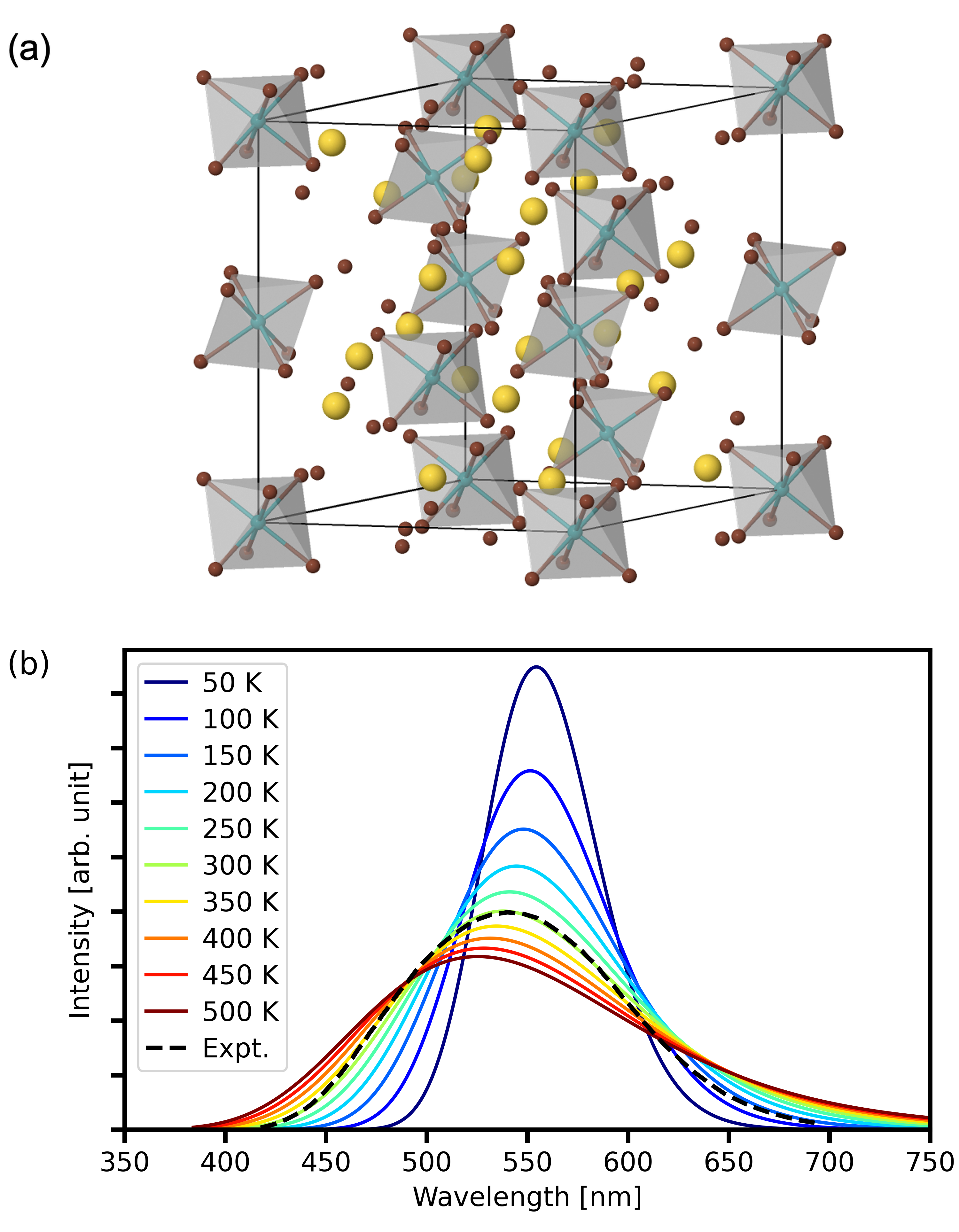}
\caption{(a) Primitive cell of the Cs$_4$SnBr$_6$ perovskite. The cesium, tin, and bromine atoms are shown in gold, teal, and brown, respectively. (b) Computed temperature-dependent emission line shapes of Cs$_4$SnBr$_6$ compared with the experimental spectrum at 300 K~\cite{perovskite_benin_2018}. The computed spectra were shifted to align the peak position with the experimental result. Adapted with permission from reference~\onlinecite{perovskite_jin_2024}. Copyright 2024, American Chemical Society.}
\label{fig:perovskite}
\end{figure}

Figure~\ref{fig:perovskite} shows the crystal structure of Cs$_4$SnBr$_6$ together with the computed temperature-dependent emission line shapes. The calculated spectrum at 300 K is in excellent agreement with experiment. Ongoing work aims to extend these studies using fully relativistic TDDFT, including SOC effects, which may be important for perovskites containing heavy elements.

\subsubsection{Optical absorption and emission of water and ice}

Understanding the interaction of light with water and ice is of fundamental importance in environmental, atmospheric, and astrophysical sciences. The WEST G$_0$W$_0$ implementation was combined with molecular dynamics simulations and ML potentials to investigate the impact of nuclear quantum effects (NQEs) on the electronic structure of water and ice~\cite{ice_berrens_2024}, revealing that NQEs induce a larger renormalization of the fundamental gap in ice than in water, leading to similar gaps in the two systems, consistent with experimental estimates. The WEST BSE implementation was employed to compute optical absorption spectra of several configurations of liquid water containing 64 molecules, extracted from molecular dynamics trajectories, as well as for a proton-disordered ice Ih model containing 96 water molecules~\cite{bse_nguyen_2019}. The calculated absorption spectra exhibit remarkable agreement with experiment, reproducing both the relative peak positions and intensities over a broad energy range. Additional calculations for a larger supercell containing 256 water molecules showed that finite-size effects, while not entirely negligible, have only a modest impact on the absorption energies.

\begin{figure}[ht!]
\includegraphics[width=0.48\textwidth]{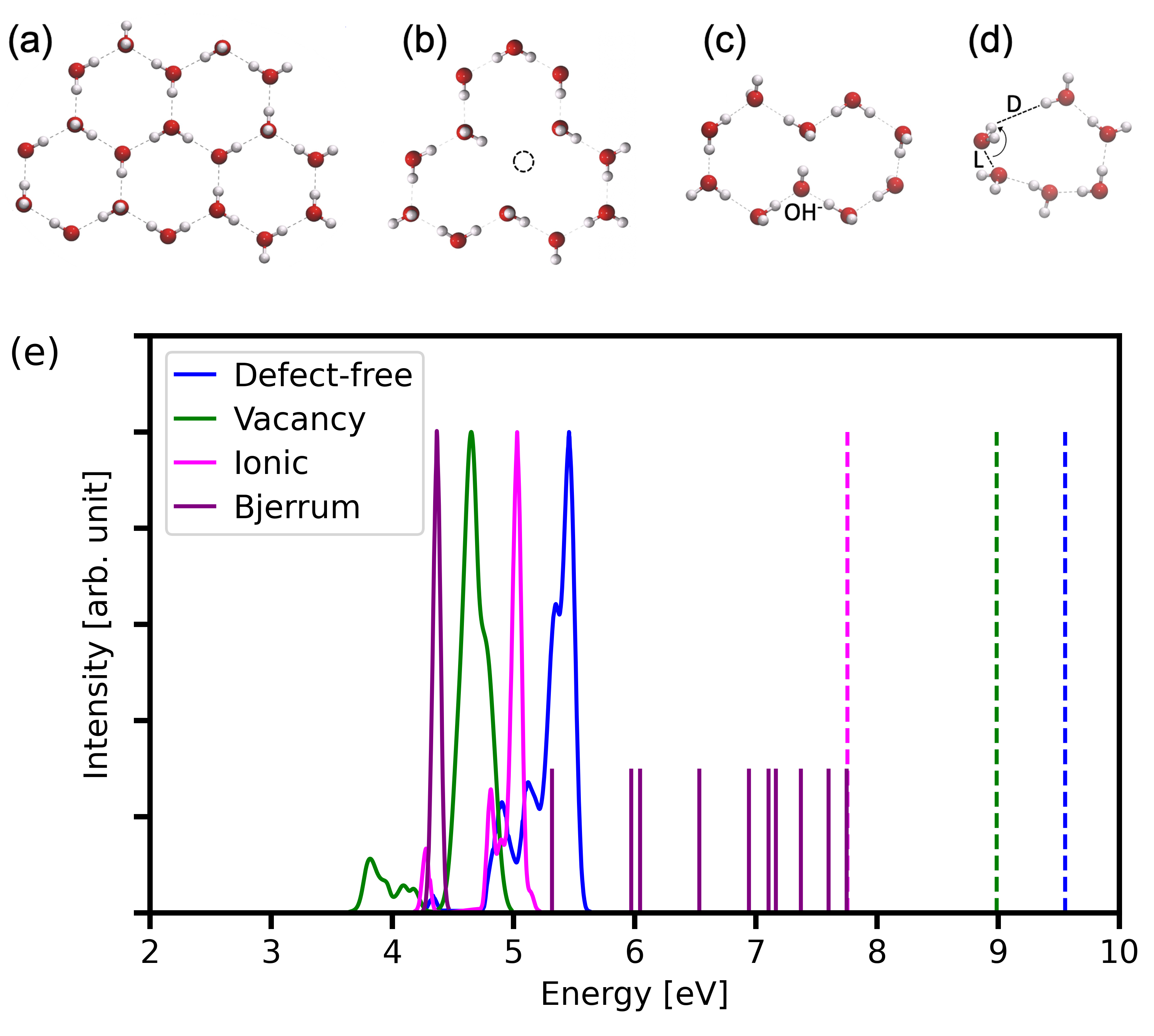}
\caption{(a) Defect-free proton-disordered ice Ih. (b) Ice containing vacancies. (c) Ice with ionic (OH$^-$) defects. (d) Ice with L- and D-type Bjerrum defects. Hydrogen and oxygen atoms are shown in white and red, respectively. (e) Computed absorption and emission onset energies for defect-free ice and ice containing vacancies, ionic defects, and Bjerrum defects. Vertical dashed lines denote the average absorption onset energies. Vertical solid lines denote the absorption onset energies of the final ten Bjerrum defect configurations obtained from nudged elastic band calculations. Colored solid curves represent Gaussian kernel density estimates fitted to the normalized histograms of the emission onset energy distributions for each model. Adapted from reference~\onlinecite{ice_monti_2025}. Copyright 2025, Monti et al.}
\label{fig:ice}
\end{figure}

More recently, the hybrid TDDFT implementation in WEST, with analytical excited-state forces, was employed to investigate how defects modify the photophysics and photochemistry of ice~\cite{ice_monti_2025}. Absorption and emission processes were studied in ice Ih containing several types of defects, including vacancies, ionic (OH$^-$) defects, and Bjerrum defects, as shown in figure~\ref{fig:ice}(a). Absorption and emission energies were averaged over multiple configurations, and for Bjerrum defects, the defect migration pathway was characterized using nudged elastic band calculations. Visualization of unrelaxed differential densities provided direct insights into the nature of the excitation processes. The computed onset absorption and emission energies for different ice models are summarized in figure~\ref{fig:ice}(b). Relative to defect-free ice, which has an average absorption onset of approximately 9.5 eV, these defects introduce gap states that red-shift the absorption onset by about 0.6 to 3 eV, depending on the defect type. The corresponding emission energies are further red-shifted, giving Stokes shifts of approximately 3 to 5 eV. Future studies using WEST will examine the role of non-radiative decay mechanisms, as well as NQEs on absorption and emission processes.

\section{Summary and outlook}
\label{sec:summary}

In this work, we presented an overview of the WEST code for large-scale excited-state simulations based on G$_0$W$_0$, QDET, BSE, and TDDFT calculations. WEST combines methodological developments, high-performance implementations, and interoperable workflows to enable accurate first-principles investigations of excited-state phenomena in complex materials. A defining feature of WEST is the elimination of explicit summations over empty electronic states across all major functionalities. By reformulating excited-state methods within the frameworks of DFPT, DMPT, and quantum embedding, WEST significantly reduces the computational and memory bottlenecks that have limited the applicability of excited-state electronic structure methods, making it feasible to carry out accurate simulations of systems containing hundreds to thousands of atoms and compute a broad range of excited-state properties and processes, including QP energies, excitation energies, optical spectra, excited-state forces, radiative and non-radiative decay rates, and quantum vibronic effects. Representative applications illustrate how these capabilities facilitate or unlock realistic simulations of complex and heterogeneous materials, including defects near dislocations and interfaces that require large supercells to capture finite-size effects and structural inhomogeneity. In addition, WEST is broadly applicable across multiple classes of materials and scientific problems, including molecules, semiconductors, oxides, nitrides, perovskites, 2D systems, water, and ice, all in experimentally relevant configurations.

The design of WEST emphasizes modularity, interoperability, and reproducibility. WEST is developed as an open-source project with publicly available source code, documentation, and tutorials. Core numerical algorithms are shared across different functionalities, enabling methodological consistency and efficient software maintenance. Through interfaces with external software packages, WEST can be integrated into automated computational workflows, promoting the rapid development of new methodologies and collaboration in diverse areas of materials science. Continued efforts towards portable programming models and hardware-agnostic optimization strategies will help ensure that the code can adapt to rapidly evolving HPC architectures. Quantum computing represents another promising direction, and WEST already supports the diagonalization of QDET effective Hamiltonians using quantum algorithms. As quantum hardware continues to mature, hybrid quantum-classical workflows may provide new pathways for treating strongly correlated systems that remain challenging for classical computers. In this context, scalable embedding approaches such as QDET are especially attractive because they naturally partition complex materials systems into correlated active spaces amenable to quantum computation while retaining an accurate first-principles description of the environment.

The combination of scalable excited-state methods, exascale and quantum computing, and AI/ML opens new opportunities for performing sustained large-scale campaigns for materials discovery. With access to massive GPU resources, one could envision generating extensive databases of excited-state properties across broad chemical and structural spaces for training and validating ML models for applications in, for example, quantum information science, spectroscopy, optoelectronics, and photocatalysis. Such datasets also accelerate the discovery of descriptors linking structures to desired properties and enable autonomous workflows combining ground- and excited-state screening and characterization that would be prohibitively expensive to investigate exhaustively using conventional approaches. More broadly, the integration of WEST with workflow managers, database infrastructures, and AI agents could enable self-driving computational frameworks capable of automated exploration of chemical and configurational spaces. At the same time, continued development of physics-based electronic structure methods remains essential. First-principles theories not only provide high-fidelity training and benchmark datasets for AI/ML models, but also deliver physical insights and interpretability that guide the refinement of data-driven approaches. We anticipate that continued developments along these directions will further expand the range of accessible systems and phenomena, enabling quantitatively predictive simulations of excited-state processes in complex materials at unprecedented scales.

\begin{acknowledgments}
The development of the WEST code has been supported by the Midwest Integrated Center for Computational Materials (MICCoM), as part of the Computational Materials Sciences Program funded by the U.S. Department of Energy, Office of Science, Basic Energy Sciences, Materials Sciences, and Engineering Division through Argonne National Laboratory. We gratefully acknowledge numerous fruitful discussions with and contributions from Nicholas Brawand, Sijia Dong, Matteo Gerosa, Fran\c{c}ois Gygi, Ikutaro Hamada, Lan Huang, He Ma, Ryan McAvoy, Ngoc Linh Nguyen, Yuan Ping, Peter Scherpelz, Nan Sheng, Jonathan Skone, Christian Vorwerk, Andrew Xu, Han Yang, and Huihuo Zheng. Computational resources have been provided by the Research Computing Center at the University of Chicago, the National Energy Research Scientific Computing Center, a U.S. Department of Energy Office of Science User Facility operated under Contract No. DE-AC02-05CH11231, and the Argonne Leadership Computing Facility at Argonne National Laboratory, a U.S. Department of Energy Office of Science User Facility operated under Contract No. DE-AC02-06CH11357. The Flatiron Institute is a division of the Simons Foundation.
\end{acknowledgments}

\section*{Data availability}

The WEST code is publicly available at \url{https://github.com/west-code-development/}. Documentation and tutorials are available at \url{https://west-code.org/}. Data sharing is not applicable to this article as no new data were created or analyzed in this study.

\bibliographystyle{apsrev4-2}
\bibliography{west}

\end{document}